\documentclass[12pt]{article}
\pdfoutput=1
\usepackage{jheppub}
\usepackage{verbatim}
\usepackage{amsmath}
\usepackage{amssymb}
\usepackage{amsthm}
\usepackage{slashed}
\usepackage{amsfonts}
\usepackage{breqn}
\usepackage{interval}
\usepackage{braket}
\usepackage[utf8]{inputenc}
\usepackage[normalem]{ulem}

\def\be{\begin{equation}}
\def\ee{\end{equation}}
\def\ba#1\ea{\begin{align}#1\end{align}}
\def\bg#1\eg{\begin{gather}#1\end{gather}}
\def\bm#1\em{\begin{multline}#1\end{multline}}
\def\bmd#1\emd{\begin{multlined}#1\end{multlined}}

\def\e{\epsilon}

\def\({\left(}
\def\){\right)}
\def\[{\left[}
\def\]{\right]}
\def\<{\langle}
\def\>{\rangle}

\newcommand{\bfig}{\begin{figure}\begin{center}}
\newcommand{\efig}{\end{center}\end{figure}}
\newcommand{\bi}{\begin{itemize}}
\newcommand{\ei}{\end{itemize}}
\newcommand{\tr}{\mathrm{tr}}

\newcommand{\Tr}{\mathrm{Tr}}

\theoremstyle{definition}

\DeclareMathOperator{\diag}{diag} 

\begin{document}

%\subheader{empty}
\title{BPS coherent states and  localization}
\author[]{David Berenstein,}
\author[]{Shannon Wang}
\affiliation[]{Department of Physics, University of California, Santa Barbara, CA 93106, USA}
\emailAdd{dberens@physics.ucsb.edu}
\emailAdd{shannonwang@physics.ucsb.edu}

\abstract{We introduce coherent states averaged over a  gauge group action to study correlators of half BPS states in ${\cal N}=4 $ SYM theory. The overlaps of these averaged coherent states are a generating function of correlators and can be written in terms of the Harish-Chandra-Itzykzon-Zuber (HCIZ) integral.  
We show that this formula immediately leads to a computation of the normalization of two point functions in terms of characters obtained originally in the work of Corley, Jevicki and Ramgoolam. We also find various generalizations for $A_{n-1}$ quivers that follow directly from other solvable integrals over unitary groups. All of these can be computed using localization methods.  When we promote the parameters of the generating function to collective coordinates, there is a dominant saddle that controls the effective action of these coherent states in the regime where they describe single AdS giant gravitons. We also discuss how to add open strings to this formulation. These will produce calculations that rely on correlators of matrix components of unitaries in the ensemble that is determined by the HCIZ integral to determine anomalous dimensions. We also discuss how sphere giants arise from Grassman integrals, how one gets a dominant saddle and how open strings are added in that case. The fact that there is a dominant saddle helps to understand how a $1/N$ 
expansion arises for open strings. 
 We generalize the coherent state idea to study $1/4$ and $1/8$ BPS states as more general integrals over unitary groups.  }

\maketitle

\section{Introduction}

There is a classic combinatorial result for two point functions of gauge invariant half BPS operators in ${\cal N}=4 $ SYM \cite{Corley:2001zk}. Let $X,Y,Z$ be the chiral adjoint scalar field of ${\cal N}=4$ SYM with respect to an ${\cal N}=1$ decomposition.
Let $R,R'$ be two different representations of $U(N)$ characterized by Young diagrams with $n$ boxes and let $\chi_R$ be the character of $U(N)$ in the corresponding representation. Then the following is true:
\begin{equation}
\braket{\chi_R(\bar Z)(x) \chi_{R'}(Z)(0)} = \frac{1}{C^n |x|^{2 n}} \delta_{R,R'} f_R, \label{eq:fR} 
\end{equation}
where $C$ is a normalization constant that depends on conventions. The quantity $f_R$ is a product over the labels of the boxes of the Young diagram associated to $R$, and is defined as:
\begin{equation}
f_R= \prod_{i,j \in \hbox{boxes}} (N+i-j),
\end{equation}
where $i$ moves to the right along the rows, $j$ moves vertically downward along the columns, and both indices start at $(i,j)=(1,1)$ in the leftmost upper corner.

Various arguments suggest that this result is not renormalized \cite{Lee:1998bxa,Baggio:2012rr} (see also \cite{Rastelli:2016nze} and references therein). 
These combinatorial calculations have been extended to other (free field) quiver setups in the works \cite{Dey:2011ea,Caputa:2012dg,Pasukonis:2013ts} (the results are written succinctly in \cite{Berenstein:2015ooa} in terms of generalized free fermions).

Part of the importance of the characters, apart from their orthogonality, is that they can also be interpreted geometrically in terms of D-branes, particularly, giant gravitons \cite{McGreevy:2000cw}. Giant gravitons expanding along the sphere directions (sphere giants for us) arise from column representations \cite{Balasubramanian:2001nh}, and giant gravitons expanding in AdS \cite{Grisaru:2000zn,Hashimoto:2000zp} (AdS giants for us) arise from large row representations \cite{Corley:2001zk}. These have served as a starting point to compute the anomalous dimensions of D-branes and the open strings ending on them. There is a combinatorial formalism developed in the works \cite{Balasubramanian:2004nb,deMelloKoch:2007rqf,deMelloKoch:2007nbd,Bekker:2007ea} to add open strings. The main issue with these approaches is that they are computationally very difficult to master; the required combinatoric calculus is laborious. We can ask if there is another way to arrive at these results that might lessen the burden of computations and provide additional intuition to the dynamics of these setups. 

When one works in less supersymmetric situations, such as with $1/4$ or $1/8$ BPS states, there are generalized orthogonal bases at zero coupling called restricted Schur bases (see \cite{deMelloKoch:2012sie,Mattioli:2016gyl} and references therein). However, as soon as one turns on the coupling constant of ${\cal N}=4$ SYM, one expects that the dynamics (at least semiclassically) reduce to some type of model of commuting matrices \cite{Berenstein:2005aa}. Such commuting matrix models are an ad-hoc uncontrolled approximation of the dynamics; they can be used to mimic the droplet picture of half BPS states in terms of free fermions in two dimensions \cite{Berenstein:2004kk} and extend the picture to more complicated setups with less supersymmetry where there is eigenvalue repulsion, but no fermions. The two dimensional droplet picture can also be seen directly in supergravity solutions \cite{Lin:2004nb}. Is there a systematic way to do calculations with these more general states that has less supersymmetry and embodies the spirit of commuting matrices, but is actually a complete field theory calculation that can be done ab initio? 

In this paper we will see that the answers to both of these questions is yes-- we can lessen the burden of computations for half BPS states (with strings attached) and find an exact {\em commuting matrix} model that captures $1/4$ and $1/8$ BPS states. 
The technique we introduce will reproduce all the results in equation \eqref{eq:fR} from a generating function. Similarly, we will discover a generating function that captures all $1/4$ and bosonic $1/8$ BPS states that survive at one loop.
In this second problem, the basis we find is implicit, rather than the explicit character basis described above.

Let us write the basic idea. The first step is to realize that when studying local operator insertions in the ${\cal N}=4 $ conformal field theory, one can equally well describe the states on the cylinder $S^3\times {\mathbb R}$ for a real quantum system, rather than the Euclidean formulation. That is, one uses radial quantization to turn the problem into quantum mechanics. Following \cite{Berenstein:2004kk}, one replaces the matrix scalar operator $Z(0) \leftrightarrow a_z^\dagger$ with the raising operator of the s-wave of the field $Z$ on the sphere including the matrix indices. Here, these indices are implicit. The free field correlators of $Z$ appearing in \eqref{eq:fR} can equally well be described by overlaps of states in the Fock space of states of the $a_Z^\dagger$ that are gauge invariant.  At this stage, we have only applied the operator state correspondence in the conformal field theory.

The next step is to think of the dynamics of the oscillators semiclassically by introducing coherent states. We start with the following object:
\begin{equation}
\ket \Lambda \sim \exp( \tr(\Lambda a^\dagger))\ket 0,\label{eq:naive}
\end{equation}
where $\Lambda$ is an $N\times N$ matrix of parameters. The trace indicates that we have a general linear combination of all possible raising operators.

Coherent states have the property that they are overcomplete. They have minimal uncertainty; they behave classically, but they are also eigenstates of the lowering operators. A formalism that can deal with these states is in principle able to deal with all the information of the Hilbert space, because of overcompleteness. The obvious problem with the object introduced in \eqref{eq:naive} is that it is not gauge invariant. We solve this issue by projecting the answer onto gauge invariant states, which we achieve by introducing an averaging over the gauge group. That is, we correct our na\"\i ve coherent state by the following:
\begin{equation}
\ket \Lambda = \frac 1{\int dU }\ \int dU \exp( \tr(U \Lambda U^{-1}  a^\dagger))\ket 0\label{eq:definition}.
\end{equation}

Because we projected an overcomplete basis to the set of gauge invariant states, we have an overcomplete basis of the gauge invariant states. 
One can check that the state defined this way is still a coherent state as far as gauge invariant combinations of lowering operators are concerned:
\begin{equation}
\tr(a_Z^n) \ket \Lambda= \tr(\Lambda^n) \ket \Lambda
\end{equation}
These expressions only depend on the eigenvalues of $\Lambda$. At this stage, we can think of $\Lambda$ as a diagonal matrix without loss of generality.

The matrix integrals that appear are well known. If $a^\dagger$ were a c-number matrix rather than a set of operators, then these would be the integrals of Harish-Chandra-Itzykson-Zuber \cite{harish1957differential,Itzykson:1979fi} (we will call this intgeral the HCIZ integral in this paper). Such integrals can be computed in a variety of ways. We refer the reader to the review paper \cite{Zinn-Justin:2002rai} (and also \cite{Morozov:2009jv} ) for a list of methods and references. We will liberally make use of the collected results in that paper. An important observation is that the HCIZ integral can be computed by localization \cite{Duistermaat:1982vw}. 
The overlaps 
\begin{equation}
I(\Lambda',\Lambda) = \braket{\Lambda'|\Lambda} \label{eq:overlap}
\end{equation}
can be computed exactly with the HCIZ integral. Upon writing equation \eqref{eq:definition} in a character expansion, we can recover all 
of the overlaps in \eqref{eq:fR} by comparing it to the character expansion of the overlap integral itself \eqref{eq:overlap}.

We are repackaging a lot of non-trivial combinatorial information in the manipulations of the coherent state object itself. The fact that the final result is an exact sum over saddles makes it possible to understand approximations to calculations that are not apparent in the combinatorial expressions that were performed to arrive at \eqref{eq:fR}. This idea extends to insertions of open strings, which we will describe in this formalism as well. The idea is to understand which saddle dominates and in what regimes. Once we have the coherent states, we can promote the $\Lambda$ eigenvalue parameters to collective coordinates and find a coherent state effective action for the parameters $\Lambda$ that describe the dynamics we are interested in. We extend this idea to $A_{n-1}$ quiver theories and to states that preserve less supersymmetry. The new idea is that for $1/4$ and $1/8$ BPS states, we need to introduce more than one matrix $\Lambda_{X,Y,Z}$. When we insist on the 1-loop anomalous dimension of these states vanishing, we find that the three matrices must commute and be able to be diagonalized simultaneously. We thus find a generalization of the HCIZ integral that satisfies some of the conditions for evaluation by localization and embodies the commuting matrix model reduction to eigenvalues. The point is that the matrices that commute are not the original fields. They are the collective coordinate parameters of the states in question.

The idea of localization in ${\cal N}=4 $ SYM is important for many other observables. In particular, Wilson loop correlators reduce to matrix model computations \cite{Drukker:2000rr,Erickson:2000af}. These are exact results, which arise from a localization argument \cite{Pestun:2007rz}; all the important computations are done with free fields. For the results leading to equation \eqref{eq:fR}, this also holds true: the computation arises from free fields. Thus in this paper, we are seeing a new application of the Harish-Chandra-Itzykson-Zuber integral; the fact that it can be described by localization arguments becomes important as we try to find approximations to the physics by looking at the dominant saddle. A general review of localization methods can be found in \cite{Pestun:2016zxk}.

The paper is organized as follows. In section \ref{sec:HO} we start with a model of a single harmonic oscillator and a gauged pair of harmonic oscillators to establish the method we will use later. The goal is to show that the denominators in a generating function of coherent states encode the information of the norms of states that are defined algebraically from the vacuum. 
Then in section \ref{sec:COH}, we introduce the main types of objects we study in this paper: coherent states in matrix models averaged over a gauge orbit. We show how to compute overlaps of these states in terms of the Harish-Chandra-Itzykson-Zuber integrals and study various generalizations of these ideas to simply laced quivers. We show that this method reproduces many results that are known in the literature. We also discuss the fact that in the integral representation, one gets exact sums over saddles. This becomes important later on when we discuss approximations of the dominant saddle and other extensions of these ideas.

 In section \ref{sec:DET} we study generating functions made by determinants rather than coherent states. These are related to sphere giant gravitons in $AdS_5\times S^5$. The point is to show that these objects admit an integral representation with a dominant saddle. The idea is to introduce fermions so that the determinant arises from Grassman integrals. Overlaps can be computed with the help of the Hubbard-Stratonovich trick and the fermions can be eliminated completely in terms of a pair of complex auxiliary variables. The integral over these variables reproduces many results. We show how these fermionic variables allow us to introduce open strings attached to the giant gravitons and demonstrate how this formalism simplifies other approaches in the literature. 
  In section \ref{sec:COL} we show how the fact that there are dominant saddles in the integrals allow one to not only promote the parameters of the gauge invariant coherent states to collective coordinates, but also calculate the effective action for them. In the HCIZ formula, these parameters are associated to multiple AdS giant gravitons. We explain how open strings are added to these configurations as well. We then turn to the problem of studying multiple sphere giant gravitons and argue that the correct multi-giant generalization involves products of determinants. This uses additional information involving character formulas and the Cauchy identity. We also explain how one has a Hilbert space of strings attached to multiple giants and explain the origin of the Gauss' law constraint. In section \ref{sec:QBP} we extend the idea of averaged coherent states to $1/4$ and $1/8$ bosonic BPS states. These require matrix parameters that commute with each other as is expected from the moduli space of vacua of these theories. We show that the saddles for half BPS states survive and focus on the dominant saddle for a single large eigenvalue and explain some of the differences that appear in the collective coordinate representation of these states. Finally, we close with a brief discussion of our results and present a possible extension of the ideas we discuss here in \ref{sec:DIS}.

\section{Warmup: the harmonic oscillator and the gauged double harmonic oscillator}\label{sec:HO}

Let us start with the simplest problem of a single harmonic oscillator. This example is intended to showcase the method we will use later in more complex settings.  
The idea is to consider a harmonic oscillator in the Hamiltonian formalism, described by a Weyl algebra constructed with a raising and a lowering operator, $[a, a^\dagger]=1$. The ground state is described by $a\ket 0=0$. Consider now the following generating function of states:
\begin{equation}
F[z]= \ket z=  \exp(z a^\dagger) \ket 0.
\end{equation}
Because $(a^\dagger)^k\ket 0$ is a complete basis of states, in principle $F(Z)$ contains (all of the) information about the full Hilbert space of states.

For the time being, the variable $z$ is a formal parameter. If we call the non-normalized state $\ket n= (a^\dagger)^n \ket 0$, we can ask how to compute its norm from $F[z]$ and indeed, the overlaps $\braket {m|n}$ for all $m,n$. Once we decide that $F[z]$ is well defined, we can think of it like a state $\ket z$ where $z$ is an actual complex variable and not just a formal parameter.
The idea is to compute the overlap:
\begin{equation}
\braket{\xi | z} = \bar F[\bar \xi] * F[z] = \bra 0 \exp(\bar \xi a) \exp(z a^\dagger) \ket 0,
\end{equation}
where $\bar F$ is the adjoint of the generating function and $\bar \xi$ is another formal parameter. We should notice that in the bra-ket notation, $(\ket \xi)^\dagger= \bra \xi$ and includes an implicit complex conjugation.
We make this explicit in $\bar F$ and implicit in $\bra \xi$. Hopefully, this will not lead to confusion.
 
There are two ways to do the calculation. First, we can expand the double series to obtain:
\begin{equation}
\braket{ \xi | z} = \sum_{m,n=0}^\infty \frac{\bar\xi^m z^n}{m! n!} \braket{m|n} .\label{eq:double}
\end{equation}
The other way to do the calculation is to contract the $a^\dagger$ and the $a$ (using the Baker-Campbell-Hausdorff formula) to obtain: 
\begin{equation}
\braket{ \xi | z} = \exp(\bar \xi z) =\sum_{n=0}^\infty  \frac{\bar\xi^n z^n}{n!}.\label{eq:eval}
\end{equation}
Comparing the two formulas, we find that the coefficient of $\bar\xi^m z^n$ for $n\neq m$ vanishes, which is to say that the states $\ket m$ and $\ket n$ are orthogonal to each other if $n\neq m$. We also find, comparing the coefficients of $\bar \xi^n z^n$, that:
\begin{equation}
\frac{\braket{n|n}}{n! n!} = \frac 1 {n!},
\end{equation}
so that $\braket{n|n} = n!$. This can be proved immediately using the raising/lowering operator algebra. The point is that $n!$ is the denominator of the terms of $\bar \xi^n z^n$ in the exponential function.

Now, because the exponential function has an infinite radius of convergence, the overlaps are well defined for any value of the complex variables $\bar \xi, z$. In particular, the norm:
\begin{equation}
\braket{z|z}= \exp(\bar z z)>0,
\end{equation}
is positive definite if $\bar z$ is the complex conjugate of $z$ and defines an $L^2$ normalizable state in the Hilbert space.

Coherent states also satisfy $a \ket z = z \ket z \simeq \partial_{a^\dagger} F[z] $, so it is easy to evaluate matrix elements of $(a^\dagger)^k a^m$ from the generating function, giving us:
\begin{equation}
\bar F[\bar \xi] * (a^\dagger)^k a^m *F[z] = z^m\bar  \xi^k \braket{\bar \xi | z},
\end{equation}
which lets us identify operationally $a^\dagger \sim \partial_z$ and $a \sim \partial_{\bar \xi}$ as far as normal ordered computations go. The point is that the generating function is not only a generating function of states, but can also be used to compute matrix elements by comparing the double expansion \eqref{eq:double} with the evaluation formula similar to \eqref{eq:eval}.  

We now go to our second example, where we have two oscillator algebras with raising operators $a^\dagger, b^\dagger$, and consider the symmetry generator $\hat Q= a^\dagger a - b^\dagger b $. We want to build a generating function as above, using gauge invariant states where $Q=0$.
A na\"\i ve guess is to do the following:
\begin{equation}
\exp( z b^\dagger a^\dagger) \ket 0.
\end{equation}

This turns out not to be optimal: the Baker-Campbell-Hausdorff trick doesn't yield a simple answer. Another option is to use a simple coherent state:
\begin{equation}
\ket{\alpha,\beta}= \exp(\alpha a^\dagger +\beta b^\dagger)\ket 0 = \sum \frac{(\alpha^m \beta^n)}{m! n!} \ket m \otimes \ket n,
\end{equation}
but we notice that the generating function also contains non gauge invariant states. We need to project them onto the $n=m$ subset. Because we start with full coherent states, we have all of the information of the Hilbert space, including the states that are not gauge invariant. If we perform the correct projection, we should retain {\em all} the information that is gauge invariant in the generating function.

This can be done if we notice that the formal parameters $\alpha$, $\beta$ can be made to transform under a $U(1)$ symmetry that tracks the charges of $a^\dagger $ and $b^\dagger$. That is, we take $\alpha \to \exp(i \theta) \alpha$ and $\beta \to \exp(-i \theta) \beta$. Then we find:
\begin{equation}
\ket{\alpha,\beta, \theta}= \exp(\alpha e^ {i\theta}  a^\dagger +\beta e^{-i\theta} b^\dagger)\ket 0 = \sum \frac{(\alpha^m \beta^n)}{m! n!} \exp(i(m-n) \theta)  \ket m \otimes \ket n,
\end{equation}
and if we seek to only obtain the states with $n=m$, we can average over $\theta$. That is, we consider a generating function of the form: 
\begin{equation}
F[\alpha, \beta] = \frac{1}{2\pi}\int_0^{2\pi} d\theta \exp(\alpha e^ {i\theta}  a^\dagger +\beta e^{-i\theta} b^\dagger)\ket 0,
\end{equation}
where we average over the group action on the operators. 
This formal functional is given by:
\begin{equation}
F[\alpha,\beta] = \frac 1{2\pi}\int d\theta \exp(i \hat Q \theta) \exp(\alpha a^\dagger+\beta b^\dagger) \ket 0,
\end{equation}
where $\hat Q$ is the charge operator defined previously.

This is almost a coherent state, except for the group projection. It is straightforward to compute the overlap:
\begin{equation}
\bar F[\bar \alpha, \bar \beta]* F[\alpha,  \beta]= \frac{1}{(2\pi)^2} \iint_0^{2\pi}  d\tilde \theta d \theta \exp\left[ \bar \alpha \alpha \exp(i(\theta-\tilde \theta)) + \bar\beta\beta\exp(i(\tilde \theta- \theta))\right]\label{eq:bes}
\end{equation}
We can now shift variables to $\theta'= \theta-\tilde \theta, \tilde \theta$, so that one group integral becomes trivial, leaving the other to be explicitly evaluated. We find that:
\begin{equation}
\bar F[\bar \alpha, \bar \beta]* F[\alpha,  \beta] = \sum_{n=0}^\infty \frac{ (\bar \alpha \alpha \bar\beta\beta)^n}{n! n!}=I_0\left(2 \sqrt{\bar \alpha \alpha \bar\beta\beta}\right)
\end{equation}
which only depends on the gauge invariant combination of parameters $\alpha\beta$ and $\bar\alpha \bar\beta$. 
It can also be written explicitly in terms of a Bessel function. At this stage we can set $\alpha=\beta$ if we want to, as they do not have an independent meaning any longer.
We also find through comparing coefficients that:
\begin{equation}
\left(\bra {n}\otimes \bra n\right)\left ( \ket m\otimes \ket m\right) = (n!)^2 \delta_{n,m} 
\end{equation}
where again, the norm of the state is the denominator in the (integrated) generating function. 

We can check that this is an eigenstate of the gauge invariant composite $a b$ operator, finding that 
\begin{equation}
a b F[\alpha, \beta] = \alpha\beta F[\alpha, \beta] .
\end{equation} 
It is this property that makes these states more convenient: they act as coherent states for the composite gauge invariant operators built from lowering operators. 
This property can be readily used to compute matrix elements. In this example, the algebra is fairly straightforward, so the calculations can be done without the generating functions.

We can do one more variation on this calculation. The idea is to use the charge $Q=a^\dagger a-b^\dagger b-k$ where $k$ is an integer. In this case, the state $\ket0\times \ket 0$ has charge $-k$ and is not gauge invariant. The gauge invariant states are $\ket{k+n}\otimes \ket n$.
In the double sum of the coherent state:
\begin{equation}
\ket{\alpha,\beta, \theta}= \exp(\alpha e^ {i\theta}  a^\dagger +\beta e^{-i\theta} b^\dagger)\ket 0 = \sum \frac{(\alpha^m \beta^n)}{m! n!} \exp(i(m-n) \theta)  \ket m \otimes \ket n,
\end{equation}
we need to project onto states where $m-n=k$. This is straightforward. We use the Fourier transform coefficients of the generating function:
\begin{equation}
\bar F[\alpha,\beta]_k=  \frac 1{2\pi}\int_0^{2\pi} d\theta \exp(-i k \theta) \exp(\alpha e^ {i\theta}  a^\dagger +\beta e^{-i\theta} b^\dagger)\ket 0
\end{equation}
so that the overlap integral is
\begin{eqnarray}
F[\bar\alpha, \bar\beta]_k*F[\alpha,\beta]_k&=&  \frac 1{2\pi}\int_0^{2\pi} d\theta \exp(-i k \theta) \exp(\bar \alpha \alpha e^ {i\theta}  +\bar\beta \beta e^{-i\theta} )\\
&=& \sum_{n=0}^\infty \frac{(\bar\alpha \alpha)^{n+k} (\bar\beta \beta)^n}{(n+k)! n!} \\
&=& (\bar\alpha \alpha)^k \left(\sqrt{\bar\alpha \alpha\bar\beta \beta}\right)^{-k/2} I_k\left(2 \sqrt{\bar\alpha \alpha\bar\beta \beta} \right)
\end{eqnarray}
Again, the norm of the fixed charge states is the denominator $(n+k)! n!$ in the sum, and the generating function can be explicitly written in terms of Bessel functions. It is a convergent power series for all $\alpha, \beta$ in the complex plane. The norm is well defined if $\alpha$ and $\bar \alpha$ transform oppositely, which they do if they are complex conjugates of each other. These are also coherent states in the sense of being an eigenvalue of the gauge invariant composite $ab$ operator with eigenvalue $\alpha\beta$.
Notice that in this case, $\alpha$ and $\beta$ appear slightly differently in the overlap. We can think of this as an anomaly. We can also take states given by $\alpha^{-k}F[\alpha,\beta]_k$, which are still coherent states; in that case, the final answer only depends on the product $\alpha\beta$, so we can take them to be equal to each other if we want to.

\section{Half BPS coherent states in ${\cal N}=4$ SYM and some generalizations }
\label{sec:COH}

We now turn to the problem of finding coherent states for the half BPS states ${\cal N}=4 $ SYM that are gauge invariant. These states are special in that they are created by a single matrix of raising operator $(a^\dagger_Z)^i_j$. 
Under the operator state correspondence, the matrix valued operator inserted at the origin is equivalent to the raising operators $Z(0)\leftrightarrow a^\dagger_Z$, where $a^\dagger$ is the raising operator for the s-wave of the field $Z$ on $S^3$, when studying ${\cal N}=4$ SYM on the cylinder \cite{Berenstein:2004kk}.  
We first consider a na\"\i ve coherent state:
\begin{equation}
F[\Lambda]= \exp( \tr(\Lambda\cdot a^\dagger_Z))\ket 0
\end{equation}
with a matrix valued $\Lambda$ set of parameters. This is a coherent state for the gauge invariant traces $\tr(a^k)$, so that 
\begin{equation}
\tr(a^k)F[\Lambda] = \tr(\Lambda^k) F[\Lambda].
\end{equation}
Since these traces generate all the gauge invariant states from the vacuum, we notice that the only information that we get from $\Lambda$ is contained in the traces of powers of $\Lambda$. This is equivalent to knowing only the eigenvalues of $\Lambda$. In that sense, most of the parameters are redundant.
We take $\Lambda$ to be diagonal in what follows.
The next problem we have to deal with is that this is not a gauge invariant state. We now introduce the $U(N)$ group action on these states and average over the group. This will look as follows:
\begin{equation}
\label{eqn:izequation}
   F[\Lambda]= \frac 1 {Vol(U(N))}\int dU\exp\left(\Tr\left(U\Lambda U^{-1}a^{\dagger}_{Z}\right)\right)\ket 0,
\end{equation}
where $dU$ is the Haar measure. The volume of $Vol(U(N)) = \int dU$ and we will call it $Vol$ for short.
For fixed $U$, the integrand will have the same coherent state properties with respect to $\Lambda, \tilde \Lambda$ if $\Lambda\rightarrow \tilde \Lambda= U\Lambda U^{-1}$ (they are related by conjugation), so that $\tilde \Lambda$ and $\Lambda$ have the same eigenvalues and traces. 
We can think either of the matrix $\Lambda$ transforming with $U$ at fixed eigenvalues, or the matrix operator $a^\dagger _Z$ transforming with $U$. In the first case, we can think of the combination $ \tr(\Lambda\cdot a^\dagger_Z)$ as being gauge invariant if both 
$\Lambda$ and $a^\dagger$ transform opposite to each other. Diagonalizing $\Lambda$ is a gauge choice and we are summing over the gauge orbit. In the second case, we may think of this as transforming $a^\dagger$ and projecting onto the gauge invariant states at fixed $\Lambda$. 
Either way, we should think of this integral as generating all of the possible half-BPS states. 

Right now, we define $\Lambda$ as an external matrix such that when we act on $F$, the lowering operators act as $a_Z \sim U\diag\left(\lambda_1, ...\lambda_N\right) U^{-1}= U \Lambda U^{-1}$ at fixed $U$. The $U$ disappear inside traces. We now wish to find the inner product of $\bar F[\bar\Lambda']$ and $F[\Lambda]$. Using Eqn.~\ref{eqn:izequation} and the Baker-Campbell-Hausdorff formula, we arrive at:

\begin{eqnarray}
\label{eqn:dropletoscillator}
    \bar F[\bar \Lambda]* F[\Lambda] &=& \frac{1}{Vol^2}\int dU^{*}dU \bra 0\exp\left(\Tr\left(U^* \bar \Lambda^{\prime}U^{* -1}a_{Z}\right)\right)\exp\left(\Tr\left(U\Lambda U^{-1}a^{\dagger}_{Z}\right)\right)\ket 0\nonumber\\
     &=&\frac{1}{Vol^2}\int dU^{*}dU\exp\left(\Tr\left( U\Lambda U^{-1}U^{* } \bar \Lambda^{\prime}U^{*-1}\right)\right))
\end{eqnarray}
Here, there is an implicit convention for transposes in  $\bar F$ for the contraction of the raising/lowering operators that lets us concatenate the matrices in the order shown. 
Any other way of doing the contraction will give a similar answer with $U^*$ either transposed or inverted in the formulas. They are all equivalent under a change of variables in the Haar measure.
Notice that the expression above depends only on the combination $U^{-1}U^{* }$ and its inverse. We can therefore call a new group variable $\tilde U=U^{-1}U^{* }$ and still keep $U$. Since the Haar measure is group invariant, at fixed $U$, we have $d\tilde U = d U^*$, which allows us to write $dU dU^* = dU d\tilde U$. The integral over $U$ can then be done -- it cancels one factor of the volume. The end result is that the integral simplifies into: 
\begin{equation}
 \bar F[\bar \Lambda]* F[\Lambda]  = \frac{1}{Vol}\int d\tilde U\exp\left(\Tr\left( \tilde U ^{-1}\Lambda \tilde U \bar \Lambda^{\prime}\right)\right)
\end{equation}
This integral is of the same type as the original definition of the coherent state that gave rise to \eqref{eqn:izequation}, but it is now also a complex analytic function of $\Lambda,\bar \Lambda$, instead of a formal state in the Hilbert space. 
This is a well known integral: the Harish-Chandra-Itzykzon-Zuber integral (HCIZ) whose value can be computed via localization \cite{Duistermaat:1982vw}. This can not be directly done in the original generating function of states because the operator matrix $(a^\dagger)^i_j$ cannot be diagonalized. 

The integral localizes to solutions of:
\begin{equation}
\Tr(\tilde U ^{-1}[\delta U,\Lambda] \tilde U \bar \Lambda^{\prime})
\end{equation}
This is equivalent to:
\begin{equation}
[\Lambda, \tilde U \bar \Lambda^{\prime}\tilde U ^{-1}]=0\label{eq:saddle}
\end{equation}
so that $\Lambda$ and $ \tilde U \Lambda'\tilde U ^{-1}$ are diagonalized simultaneously. This means that the labels of diagonal components $\lambda_i$ and $\lambda'_i$ differ by a permutation $\sigma$. We can take the matrices to be diagonal and described by $\lambda_i, \lambda'_{\sigma(i)}$, which is to say that $U$ is a permutation matrix. The correct space for matrices $U$ is $U(N)/U(1)^N$, where the $U(1)^N$ can be taken as the matrices that commute with $\Lambda$ automatically.
The saddle value of the integrand is $\vec \lambda \cdot \vec \lambda'_{\sigma}= \sum \lambda_i \lambda'_{\sigma(i)}$. 

We know from \cite{Morozov:2009jv} that we can expand our integrand from Eq.~\ref{eqn:izequation} through a character expansion, giving a formula of the type:
\begin{equation}
F[\Lambda]= \sum_R \frac{1}{f_R} \chi_R(\Lambda) \chi_R(a^\dagger_Z)\ket 0,
\end{equation}
where we have a denominator $f_R$ that we will compute later. The denominator is found in equation (34) of  \cite{Morozov:2009jv}, or the denominator in equation (2.11b)  of \cite{Zinn-Justin:2002rai} if we divide by the prefactor of the equation. 
We will recompute the answer by using the exact evaluation of the integral.

We can do the same with the explicit HCIZ integral:
\begin{equation}
\bar F[\bar \Lambda]* F[\Lambda] = \sum_R \frac{1}{f_R} \chi_R(\bar \Lambda) \chi_R(\Lambda)
\end{equation}
Comparing coefficients of the characters of the matrices $\Lambda$ to the double sum 
\begin{equation}
\bar F[\bar \Lambda]* F[\Lambda] = \sum_{R,R'}\frac{ 1}{ f_{R}f_{R'}} \chi_{R'} (\bar \Lambda) \chi_R(\Lambda)\braket{0|\chi_{R'}(a) \chi_R(a^\dagger)|0},
\end{equation}
we arrive at $\braket{0|\chi_{R'}(a) \chi_R(a^\dagger) |0}=0$ if $R\neq R'$, and we also find: 
\begin{equation}
\frac{\braket{0|\chi_{R}(a) \chi_R(a^\dagger) |0}}{f_R^2}= \frac 1{f_R}
\end{equation}

That is, the characters are orthogonal to each other, and the norm of each of the characters is the denominator $f_R$. This should be contrasted with the explicit combinatorial derivation in \cite{Corley:2001zk}. The reader can check that the answer quoted above and the result
of the combinatorial formula match each other.

Now, we will compute $f_R$ directly. This is something that can be done directly from the evaluation of the HCIZ integral. The first step is to understand that the representations appearing in the equation are labeled by Young diagrams for $U(N)$. Each diagram is characterized by the length of the rows, which appear in descending order $j_1\geq j_2 \geq \dots j_N$.

We need the explicit Weyl character formula:
\begin{equation}
    \chi_{j_i}(\Lambda) = \frac{\det\left(\lambda_k^{j_i+N-i}\right)}{\Delta(\Lambda)},
\end{equation}
which is written as a ratio of determinants, where $\Delta(\Lambda)$ is the Vandermonde determinant of $\Lambda$ (we can also obtain it from the numerator by setting $j_i=0$ for all $i$). 

The second item we need is the explicit value of the HCIZ integral:
\begin{equation}
 I(\Lambda, \bar \Lambda)= \frac{1}{Vol}\int d\tilde U\exp\left(\Tr\left( \tilde U ^{-1}\Lambda \tilde U \bar \Lambda^{\prime}\right)\right) = \Omega \frac{\det\left(\exp(\lambda_i \bar\lambda' _j)\right)}{\Delta(\Lambda)\Delta(\bar \Lambda')  },
\end{equation}
where $\Omega$ is a normalization constant. The determinant in the numerator is a sum over permutations, it is an explicit sum over all the $N!$ possible saddles that are solutions of equations \eqref{eq:saddle}.

We can find $f_R$ by first multiplying the result by the product of the Vandermonde determinants. This way, we obtain:
\begin{equation}
 I(\Lambda, \bar \Lambda) \Delta(\Lambda)\Delta(\bar \Lambda')  = \Omega\det\left(\exp(\lambda_i \bar\lambda' _j)\right)= \sum_{\vec \j} \frac 1 {f_{\vec \j}}  \det\left(\lambda_k^{j_i+N-i}\right)\det\left(\bar\lambda_k^{\prime j_i+N-i}\right)\label{eq:CHIZsaddle},
\end{equation}
where we are now labeling the representations $R$ by the vector of values $\vec \j$ determining the Young diagram. 
We will consider on the right hand side the monomials of the type $\lambda_1^{r_1} \dots \lambda_N^{r_N}$ with $r_1>r_2>\dots> r_N$. These monomials are in one to one correspondence with the characters.
This corresponds to the unique term in the numerator of the determinant that is the product of the diagonal entries:
\begin{equation}
 \det\left(\lambda_k^{j_i+N-i}\right)\rightarrow \prod_i \lambda_i^{j_i+N-i}+ \dots
\end{equation}
Now, we expand the exponentials in the determinant $\det\left(\exp(\lambda_i \bar\lambda' _j)\right)$ of the evaluated HCIZ integral and use the multilinearity of the determinant to arrive at:
\begin{equation}
\det\left(\exp(\lambda_i \bar\lambda' _j)\right)= \sum_{[n]} \frac 1{[ n]!} \det ((\lambda_i \bar\lambda' _j)^{n_i}) = \sum_{[n]} \frac 1{[ n]!} \det (\bar\lambda _j^{\prime n_i})\prod_i \lambda_i^{n_i}\label{eq:series1}
\end{equation}
where $[n]$ is the multi-index $n_1, \dots,n_N$, while $[n]!$ is the product $\prod _jn_j!$. Restricting to the monomials with the correct descending order forces us to take $n_1>n_2\dots $ in the sum. We see that we get an explicit sum over characters if we set $n_i = j_i+N-i$, which also have this descending value property. 
We therefore find that the denominators can be readily computed:
\begin{equation}
f_{\vec \j}= [ n]! \Omega^{-1} = \Omega^{-1} \prod_i (j_i+N-i)!,
\end{equation}
Setting $f_{\vec 0}=1$ by using $\braket{0|0}=1$, we find $\Omega= \prod_{i=1}^N (N-i)! $. 

This is the same answer that was obtained by direct combinatorial methods.
Similar localization formulas exist for other groups \cite{Duistermaat:1982vw}. What is less straightforward is the corresponding character expansion. This is explained cursorily in \cite{Zinn-Justin:2002rai}. 
The goal would be to reproduce the combinatorial formulas in \cite{Caputa:2013hr,Caputa:2013vla} and check if the bases agree. 
We will not pursue this calculation in this paper. Instead, we will look at other integrals for simply laced quiver theories ($A_{n-1}$ quivers) to show how the method works in those cases as well.

Let us start with a gauge theory of $U(N_1)\times U(N_2)$ matrices (we start with $N_1=N_2$) and consider a pair of bifundamental fields $a^\dagger_{12},  a^\dagger_{21}$ in the $(\bar N_1,N_2)$ and the  $(N_2, \bar N_1)$ representations.
We want to build the same type of coherent states as above. We start with: 
\begin{equation}
F[\Lambda]\sim\exp\left( \Tr(\Lambda_{21}  \cdot a^\dagger_{12} + \Lambda_{12}\cdot a^\dagger_{21} )\right)\ket 0
\end{equation}
The idea is that $\Lambda_{ij}$ is in the dual space of $a^\dagger_{ji}$, so we reverse the order of the $i,j$ labels. In this sense, the lowering operators are also reversed $a_{ji}= (a^\dagger_{ij})^\dagger$. 
Just like before, we need to average over the group $U(N_1)\times U(N_2)$. This is done by the following procedure:
\begin{equation}
F[\Lambda]= \frac1{\prod_i Vol_i}\int \prod_{i=1}^2 dU_i\exp\left( \Tr(\Lambda_{21} U_1  a^\dagger_{12} U_2^{-1}+ \Lambda_{12}U_2 a^\dagger_{21}U_1^{-1} \right)\ket 0
\end{equation}
where all the contractions are matrix multiplications.

With the usual use of the Baker-Campbell-Hausdorff formula, we arrive at a formula where we end up replacing $a^\dagger_{12}$ by the $\bar \Lambda'_{12}$ matrix. We obtain: 
\begin{equation}
\bar F[\bar \Lambda]*F[\Lambda]= \frac1{\prod_i Vol_i}\int \prod_{i=1}^2 dU_i\exp\left( \Tr(\Lambda_{21} U_1  \bar\Lambda'_{12} U_2^{-1}+ \Lambda_{12}U_2 \bar \Lambda'_{21} U_1^{-1}) \right)
\end{equation}
These are well known generalizations of the HCIZ integral, first solved in \cite{Jackson:1996jb}. We now assume, without loss of generality, that the matrices $\Lambda, \bar \Lambda'$ are diagonal. The eigenvalues are:
\begin{eqnarray}
\Lambda_{12} \equiv \diag( \lambda_{(12)}^{1} \dots \lambda_{(12)}^{\min(N_1, N_2)})\\
\bar \Lambda_{12} \equiv \diag(\bar\lambda_{(12)}^{1} \dots \bar\lambda_{(12)}^{\min(N_1, N_2)})
\end{eqnarray}
Here, the $(ij)$ label of the matrix is in the lower component, and the upper components label the eigenvalues. If $N_1\neq N_2$, the matrix has diagonal entries to the extent that it is permitted, and the off-diagonal elements vanish.
Let $A^\alpha = \prod_{ij}   \lambda_{(ij)}^{\alpha}$ be the diagonal product of the $\Lambda$ matrices. Similarly $\bar A^{\beta} = \prod_{ij}   \lambda_{(ij)}^{\beta}$.  The integral can also be done with localization methods.
First, when $N_1= N_2$, we need to be careful over what we mean by localization. The original localization formula for the HCIZ integral is evaluated on a compact complex manifold $U(N)/U(1)^N$. It is important that we do the same here. However, we are doing integrals over the full $U(N_i)$ and not just $U(N_i)/U(1)^{N_i}$. We need to separate the $U(1)^{N_i}$ explicitly. These are tori, so we have variables $\exp(i \theta_i^\alpha)$ for each group. We choose these to be multiplying $U$ on the left. 
We define $U_1 = \diag(i \theta_1^\alpha) * \tilde U_1$ where $U_1^* \in U(N_1)/U(1)^{N_1}$, which is a proper complex space on which localization can be had.

When we do so, we get a similar equation to \eqref{eq:saddle}; this time, however, we need to restrict ourselves to $\delta U$ that are strictly off-diagonal (they need to be orthogonal to the tori that we selected). The result is the same: $\Lambda$ and a conjugate of $\bar\Lambda$ are mutually diagonal for all $\Lambda$. This gives rise to a common permutation of the $\lambda$ variables. 
The value of the ``action" at the saddle is:
\begin{equation}
S= \sum_\alpha \lambda_{ij} ^\alpha \bar \lambda_{ji}^{\sigma(\alpha)} \exp(i \theta_i^\alpha-i\theta_j^\alpha),  
\end{equation}
which still depends explicitly on the angles $\theta_i$. The localization integral is over the off-diagonal pieces. We still need to do an integral over the phases $\theta^{\alpha} _i$.  We may simplify the integral by absorbing the phases into the $\lambda_i$.
The denominator is computed using the method of images, under the assumption that $S$ is real.
The result is:
\begin{equation}
\sqrt{\det(\delta_U^2 S)} = \Delta(A)\Delta(\bar A),
\end{equation}
a product of Vandermonde determinants.
The localization formula (for each saddle) we need is then given by the following integral:
\begin{equation}
I_{sad}=\frac 1{\Delta(A)\Delta(\bar A)}\int \prod d\theta \exp\left[\sum_{\alpha, i} \lambda_{ij} ^\alpha \bar \lambda_{ji}^{\sigma(\alpha)} \exp(i \theta_i^\alpha-i\theta_j^\alpha)  \right]\label{eq:gensad}
\end{equation}
The integral is done by expanding the exponential as a series and using the binomial expansion. 
We find:
 \begin{equation}
\frac 1{\Delta(A)\Delta(\bar A)}\sum_{[n]} \frac{1}{ [n]!^2} A^{[n]} \bar A_\sigma^{[n]} 
\end{equation}
This can be understood as $N$ copies of the computation in \eqref{eq:bes}, so we get a product of Bessel functions. 
Each saddle also has a sign, $(-1)^\sigma$. We find that up to a normalization constant $\Omega$, we have:
\begin{equation}
\bar F[\bar \Lambda]*F[\Lambda]=\frac  \Omega {\Delta(A)\Delta(\bar A)}\det\left(J_0\left[2 \sqrt {( A^\alpha \bar A^\beta)}\right]\right)\label{eq:detpre}
\end{equation}
Because of the determinant structure, it admits a character expansion. This is another way to arrive at the answer by a direct computation. Again, it is the denominator of the characters that count. 
We replace $[n]!\rightarrow ([n]!)^2$ in all the formulas  in \eqref{eq:series1}. 
\begin{equation}
f_{\vec \j}= [ n]!^2 \Omega^{-1} = \Omega^{-1} \prod_k (j_k+N-k)!^2,
\end{equation}
including the normalization factor $\Omega$, which is determined by $f_{\vec 0}=1$. 

Once we go to more general $A_{n-1}$ quivers, with all groups of the same rank, the integrals that need to be done are of the type \eqref{eq:gensad}, but with the sum over $i$ containing more terms, as many as there are in the $A_{n-1}$ quivers. The most general formulas obtained this way contain different ranks for the different groups:
\begin{equation}
f_{\vec \j}= \Omega^{-1} \prod_{i,k} (j_k+N_i-k)!,
\end{equation}
again normalized to $f_{\vec 0}=0$ (see \cite{Zinn-Justin:2002rai,Morozov:2009jv}). The Bessel function gets replaced by a generalized hypergeometric series, given by:
\begin{equation}
\Phi(A\bar A) = \sum_{m} \frac{1}{\prod_i(m+N_i-N_0)!}( A^\alpha \bar A^{\sigma(\alpha)} )^m
\end{equation}
where $N_0= \min(N_i)$ and the determinant that generalizes equation \eqref{eq:detpre} is a determinant of a $N_0\times N_0$ matrix, so that the overlap reads: 
\begin{equation}
\bar F[\bar \Lambda]*F[\Lambda]=\frac  \Omega {\Delta(A)\Delta(\bar A)}\det\left(\Phi\left[ {( A^\alpha \bar A^\beta)}\right]\right)\label{eq:detpre2}
\end{equation}

When the ranks of the groups are not the same, the localization integral actually does not work. Let us explain this in the simplest setup where we use a $U(N)\times U(1)$ quiver.
The integral we need to do is:
\begin{equation}
I = \frac 1{2\pi} \frac 1{Vol}\int dU d\phi \exp\left(   {\bar a} \cdot U \cdot \ a \exp(i\phi) + \exp(-i\phi)  b U^{-1} \bar  b \right),
\end{equation}
where $a$ is a column vector, and $b$ is a row vector, given explicitly by: 
\begin{equation}
\vec a = \begin{pmatrix} \tilde a\\ 0\\ \vdots \end{pmatrix},\quad  \vec b= (\tilde b, 0 \dots 0),
\end{equation}
where the product $\vec b \cdot \vec a = \tilde a \tilde b$ is gauge invariant. 
Notice that the vectors $\vec a$ and $\vec b$ are invariant under a common $U(N-1)$ group. Thus, when we do the integral, we should do the integral over the quotient space: 
\begin{equation}
S^{2N-1}  \sim U(N)/U(N-1),
\end{equation}
which is a round sphere of dimension $2N-1$. This is not a complex manifold, but the quotient ${\mathbb CP}^N= S^{2N-1}/U(1)$ is such a space. This complex geometry would be the one where localization would take place. Instead of that, let us choose the metric of the sphere as follows:
\begin{equation}
ds^2 = \cos^2(\theta) d \phi_1^2 + d\theta^2 + \sin^2(\theta) d\Omega^2_{2N-3}  \label{eq:metric}
\end{equation}
The action has a pair of rotated vectors dotted into another such vector. This inner product has a $\cos(\theta) \exp(i \phi_1)$ factor in it. We find that: 
\begin{equation}
I \sim \int  \sin(\theta)^{2N-3} \cos(\theta) d\theta  d\phi_{12} \exp\left(   \bar {\tilde a} \tilde  a \cos(\theta) \exp(i\phi_{12}) +  \tilde b \cos(\theta) \bar {\tilde b} \exp(-i\phi_{12}) \right)
\end{equation}
where the angle $\theta\in [0,\pi/2]$ and $\phi_{12}$ is a relative angle. We can get the same type of answers if we use Euler angle parameterizations in $U(2)$ and $U(3)$ (see \cite{Byrd:1997uq,Tilma:2002ke} to get Euler angles for $SU(3)$ and $SU(N)$). 

We can do the integral explicitly in two different orders. We can expand the exponential series and pick the Fourier terms that have vanishing momentum. We obtain:
\begin{equation}
I \sim \sum_n  \int  \sin(\theta)^{2N-3} \cos(\theta) d\theta   (\bar {\tilde a} \tilde  a  \tilde b\bar {\tilde b})^n \frac 1{(2n)!}{2n \choose n} \cos(\theta)^{2n}  
\end{equation} 
The individual integrals can be written in terms of $\Gamma$ functions. Up to a normalization factor, we get:
\begin{equation}
I \sim \sum_n    (\bar {\tilde a} \tilde  a  \tilde b\bar {\tilde b})^n \frac 1{n!n!} \frac{\Gamma[N-1] \Gamma[n+1]}{\Gamma[n+N]} =  \sum_n    (\bar {\tilde a} \tilde  a  \tilde b\bar {\tilde b})^n \frac 1{n!} \frac{\Gamma[N-1]}{(n+N-1)!}
\end{equation} 
The denominator takes the form we expect: $f_n = (n+1-N_0) ! (n+N-N_0)!$, where $N_0=1$. Everything is fixed if we require that the leading term of the series is equal to one.
The second way to do the integral is to introduce a new variable $x= \cos(\theta)$ and let $y=  \bar {\tilde a} \tilde  a  \exp(i\phi_{12})+\tilde b \bar {\tilde b} \exp(-i\phi_{12})$. Then the integral takes the form:
\begin{equation}
I \propto \int d\phi\int_0^1 d x (1-x^2)^{N-2} x \exp ( x y)  
\end{equation}
The first few of the answers, for $N=2,3$ are:
\begin{eqnarray}
I_2 &=&  \int d\phi\left( \frac 1 {y^2} +\frac{e^y(y-1)}{y^2} \right)\\
I_3&=& \int d\phi y^{-4} (-6 + y^2 + 2 e^y (3 - 3 y + y^2))
\end{eqnarray}
and this suggests that there are two saddles: one at $\theta=0$ and the other at $\theta=\pi/2$. The first saddle has action $y$, and the other has action $0$. 
The measure factor from the saddle should be the maximum inverse power of $y$ in the expression. But we notice that there are curious factors of $y$ in the numerator. This is because the two endpoints of the $\theta$ integral correspond to manifolds of different dimensions. For $\theta=0$ in \eqref{eq:metric} we get a circle parametrized by $\phi_1$, whereas for $\theta=\pi/2$, we get a sphere of dimension $2N-3$. One of the two ``critical points" in $S^{2N-3}/U(1)$, if we can call them like that, is not isolated and the other one is. 
In that sense, a naive notion of localization fails. The Duistermaat-Heckman theorem requires isolated critical points. In the case of $N=2$, both of the critical points lead to circles shrinking to zero size, but these circles are not the same circles. The theorem of localization only pertains to the fixed points under the same $U(1)$ action. 
In spite of this, because we can do the integral in the other order (where we integrate the angle variables $\phi$ first), we get an expression that is a quotient of determinants that admits a character expansion. 
That is enough to show the orthogonality of character wave functions and compute the norm in terms of a denominator that fits the description above. 
Since the expression looks like sums over saddles with denominators, we can abuse the language of localization if need be.

For conformal field theories in four dimensions, we usually find ourselves in cases where all the $N_i$ are equal to each other, so the process of localization is valid. 

\section{Determinants and strings attached to them}\label{sec:DET}

Consider the following gauge invariant object:
\begin{equation}
G[\lambda]=\det( \lambda- a^\dagger)\ket 0,
\end{equation}
where $\lambda$ is a c-number formal variable. This expression can be expanded in characters (more precisely, subdeterminants) as follows:
\begin{equation}
G[\lambda]=\sum_{n=0}^N (-\lambda^n) sdet_{N-n}(a^\dagger)
\end{equation}
We can consider the overlap:
\begin{equation}
\bar G[\bar \lambda]* W[\lambda]
\end{equation}
Our goal right now is to find an expression of this overlap that can be computed with saddle point methods, as a saddle of a specific integral.
The idea is to write the determinant as a fermionic integral (we follow the setup \cite{Jiang:2019xdz,Jiang:2019zig,Chen:2019kgc,Yang:2021kot}, see also \cite{Budzik:2021fyh}): 
\begin{equation}
\det ( \lambda- a^\dagger) = \iint d\bar \xi d\xi \exp( \bar \xi ( \lambda - a^\dagger) \xi),
\end{equation}
where the fermions $\xi, \bar \xi$ are column and row vectors of size $N$. The determinant is the result of a fermion integral over an auxiliary set of fermions that can be taken to transform under $U(N)$ as a fundamental or antifundamental.
Notice that the term in the exponential is again linear in $a^\dagger$. What this means is that we may apply the Baker-Campbell-Hausdorff trick again in the overlap computation. The overlap we need is of the form:
\begin{equation}
\bar G[\bar \lambda]* G[\lambda] = \iint d\bar \chi d \chi d\bar \xi d\xi \exp( \bar \xi \lambda \xi + \bar \chi \bar \lambda \chi - \bar \chi \xi \bar \xi\chi),
\end{equation}
where the minus sign of the quartic term comes from a fermion sign.  

As is standard, we use the Hubbard-Stratonovich trick by inserting a complex boson Gaussian integral to find:
\begin{equation}
\bar G[\bar \lambda]* G[\lambda] = \iint d\bar \chi d \chi d\bar \xi d\xi d \bar\phi d\phi \exp( \bar \xi \lambda \xi + \bar \chi \bar \lambda \chi - \bar \phi \phi +i \phi \bar \xi\chi+ i \bar \phi\bar \chi \xi)
\end{equation}
The fermion integral is now diagonal in the $U(N)$ indices, so we get that:
\begin{eqnarray}
\bar G[\bar \lambda]* G[\lambda] &=& \iint d \bar\phi d\phi \exp( - \bar \phi \phi  ) \det\begin{pmatrix} \lambda & i\phi\\
i\bar \phi& \bar\lambda \end{pmatrix}^N\\
&=&  \iint d \bar\phi d\phi (\bar \lambda\lambda+\bar\phi\phi)^N \exp(-\bar\phi\phi)
\end{eqnarray}
We can now do the integral by expanding the polynomial in $\bar \lambda \lambda$ using the binomial theorem, finding that the integral over $\bar\phi\phi$ can be expressed in terms of of $\Gamma$ functions:
\begin{equation}
\bar G[\bar \lambda]* G[\lambda] = \Omega\sum_{k=0}^N {N \choose k}(\bar \lambda\lambda)^k\Gamma[N-k+1] = \Omega N! \sum_{k=0}^N \frac{(\bar \lambda\lambda)^k} {k!} \label{eq:gensersdet}
\end{equation}
up to a normalization constant of $\Omega$, which has been implicit in the measure of the integrals.
We need to normalize the answer so that the term with $(\bar\lambda \lambda)^N$ (the vacuum overlap) has a coefficient equal to one. This means that $\Omega=1$. 
We find in a straightforward manner that the subdeterminants are orthogonal and that their norm is $|sdet_{N-K}|^2 = N!/k!$; we can compare this expression to the results in \cite{Balasubramanian:2001nh,Berenstein:2013md}.
 
The results presented here are very direct. We notice that because we have an integral expression, we can evaluate it using a saddle point approximation by varying over $r= \bar \phi\phi$. The saddle is the minimum of:
\begin{equation}
-\bar\phi \phi + N \log(\bar\lambda \lambda +\bar \phi \phi) 
\end{equation}
Equivalently, using $r\equiv \bar \phi \phi$:
\begin{equation}
-1 +\frac{N}{ r +\bar \lambda\lambda}=0\label{eq:sad},
\end{equation}
we find that $r= N- \bar \lambda \lambda$. 
For the saddle point method to be a good approximation, we need the saddle to be close to the positive real axis, which is the line over which we integrate $r$. When $\bar \lambda$ and $\lambda$ are complex conjugates of each other, this requires that:
\begin{equation}
\bar\lambda\lambda <N\label{eq:range}
\end{equation}
In this setup, $\bar\phi\phi$ is of order $N$. Necessarily, so is $\lambda$, which we think of as a parameter.

We find:
\begin{equation}
\bar G[\bar \lambda]* G[\lambda] \sim \Omega' \exp(-r) \simeq N! \exp(\bar \lambda \lambda)
\end{equation} 
If we compare this expression to \eqref{eq:gensersdet}, we see that the exact answer is a truncated exponential, and that the saddle point approximation gives the exponential function. This fact was first seen in \cite{Berenstein:2013md}, but not as a saddle point with respect to
the integral representation.  The saddle makes it clear that we have a $1/N$ expansion, because of the specific  $N$ dependence of the logarithm. This is induced when we integrate out the fermions.

\subsection{ Adding open strings}

The idea of the saddle point calculation is that in the end, $\bar \lambda$ and $\lambda$ become complex variables. The generating function $G[\lambda]$ is to be considered as a state in the Hilbert space of states.
We can build other states around this state. Consider a collection of words $W_j$ made of raising operators different from $a^\dagger$ (like the ones that appear in spin chains of ${\cal N}=4$ SYM).
We can consider more general states that are of the form:
\begin{equation}
G[\lambda, W] = \int d\bar\xi d\xi \exp( \bar \xi ( \lambda - a^\dagger) \xi) \prod_j (\bar \xi W_j \xi) \ket 0.
\end{equation} 

These are open words $W_j$ with fermion flavors attached at the boundaries. The boundary fermions on the words make these states gauge invariant as well: the $\bar \xi, \xi$ transform under $U(N)$. 
In this formalism, the introduction of the fermion variables suggests that there are emergent degrees of freedom in the determinant. When we write the determinant as an integral over the fermion variables, we ``integrate out" these degrees of freedom when we do the integral. This has been studied in \cite{Budzik:2021fyh,Gaiotto:2021xce}.

Keeping the fermions in more places than just the determinant allows us to affix strings to the defect. Hence, the determinants behave like D-branes; indeed, they are supposed to be sphere giant gravitons. 
Without the explicit fermions, one would get a formalism similar to \cite{Berenstein:2013md}, which is more cumbersome. 
Attaching strings combinatorially for single determinant branes was pioneered in \cite{Berenstein:2002ke,Balasubramanian:2002sa,Berenstein:2003ah}.
This formalism with fermions does the same work more economically and moreover has a well defined saddle, which allows one to make approximations useful for computations.

Let us now show, mimicking \cite{Berenstein:2013md}, that any such $W$ should not begin or end in the letter $a^\dagger$, because we would be overcounting. Consider $W = W'a^\dagger $. The idea is to write $a^\dagger = \lambda + (a^\dagger -\lambda)$.
The term with just $\lambda$ is a c-number, so it can be taken out and written in terms of shorter words, in this case $W'$. There is a second term, which can be written as a fermion derivative as follows:
\begin{equation}
G[\lambda, W'(a^\dagger-\lambda) ] = -\int d\bar\xi d\xi \bar \xi W' \partial_{\bar \xi} \exp( \bar \xi ( \lambda - a^\dagger) \xi) \ket 0
\end{equation} 
Now we integrate the fermion derivative by parts and find that:
\begin{equation}
G[\lambda, W'(a^\dagger-\lambda) ] = -\int d\bar\xi d\xi \exp( \bar \xi ( \lambda - a^\dagger) \xi) \tr(W')\ket 0,
\end{equation} 
which is usually interpreted as the object $G$ with a closed string $\tr(W')$. These manipulations are straightforward, whereas the original combinatorial calculation was more challenging. The original combinatorial setup with the $\lambda$ acting as collective coordinates allows one to understand the boundary conditions for the closed spin chain in more detail \cite{Berenstein:2005vf,Berenstein:2005fa,Berenstein:2006qk} and is useful at higher loop orders \cite{Berenstein:2013eya,Berenstein:2014isa,Dzienkowski:2015zba}, but it becomes prohibitive to understand how the various diagram contribute at various orders in $1/N$. The saddle approximation and the introduction of fermions help facilitate the latter goal. 
The fermion variables make it easy to generalize further beyond one string and should prove helpful to understanding how strings 
split and join more generally once non-planar interactions are added. These types of words with fermions can be generated by interactions; the Hamiltonian lowering operators can bring down powers of the $\xi$ fermions. The integration by parts can also act on other insertions of $\bar \xi$, producing the splitting and joining of words.
Computing overlaps of states with strings will involve fermion correlators (this is how computations are done in \cite{Yang:2021kot}, for example). These are easy to compute at the saddle point obtained at \eqref{eq:saddle}, and involve the $2\times 2$ inverse of the quadratic form appearing in the fermion integral. Namely, we have a Feynman rule for a fermion propagator that eliminates the fermion insertions in the strings. The fermion propagator is:
\begin{equation}
\pi=
\begin{pmatrix} \lambda & i\phi\\
i\bar \phi& \bar\lambda \end{pmatrix}^{-1}
\end{equation}
Exploring these issues in detail is beyond the scope of the present paper.

\section{Collective coordinates}\label{sec:COL}

So far, we have defined coherent states labeled by either eigenvalue parameters $\Lambda\sim (\lambda^1, \dots \lambda^N)$, or by a generating function made of a single parameter $\lambda$ in the case of determinants.  
In all of these cases we have found that the overlaps of these states with different parameters can be well described by saddle points of an integral. When the integral is done by localization, the expansion in terms of a sum over a finite number of saddles is exact. When the states have the same parameters (that is, when we are computing the norm of a state), there is usually a single saddle that dominates, as we will describe. In such cases, the physics can become semiclassical.
The parameters $\lambda^i$ that describe the individual state can be promoted to collective coordinates, which will allow us to describe the dynamics of the state in terms of a simplified dynamics of the $\lambda$ parameters as functions of time. We will demonstrate this process in this section.

So far, we have not discussed the Hamiltonian of the system. We have described raising and lowering operators in a harmonic oscillator context, but at no point did we make it explicit that these are solutions of the dynamics of a quantum system with a Hamiltonian.
We want to understand how to do this directly from the $\lambda$ parameters. Rather than solve the oscillator dynamics of  $a^\dagger$ and port it over to the $\lambda$, we want to have an effective action for the $\lambda$ itself that reproduces it. The reason for doing this is that eventually the dynamics of BPS states get corrected when we add other oscillators. In that sense, we get an effective action of collective coordinates and additional excitations, which interact with one another. These interactions lead to corrections of motion in $\lambda$, but the fact that the saddles are in some sense strong saddles means that these can still be treated semiclassically and the states will have big overlaps with the coherent states described so far.

The effective action of the collective coordinates on their own is usually written as a first order formulation as follows:
\begin{equation}
S=\int dt \bra \lambda i \partial_t \ket \lambda - \braket{\lambda| \hat H| \lambda},
\end{equation}
where the first term is a Berry phase. The states $\ket \lambda$ are required to be normalized. Applying the variational principle to the action produces an approximation to the Schr\"odinger equation, restricted to the states of the prescribed form. 
Our idea is to use the saddle point expressions directly to compute the effective action $S$. This is very similar to what was done in \cite{Berenstein:2013md} for a single sphere giant graviton.

The main idea behind computing $\hat H\ket {\lambda}$ is that the energy in the generating series is equal to the number of raising operators in the expansion of the exponential. This is identical to counting powers of $\Lambda$. In that sense, we take the un-normalized $\ket{\vec\lambda}$, which is strictly holomorphic, and find that:
\begin{equation}
\hat H \ket \lambda = \sum \lambda^i \partial_{\lambda_i} \ket \lambda
\end{equation}
After this evaluation, we can rescale the state by multiplying by a c-number (the square root of the norm of the state). If we define the normalization constant as:
\begin{equation}
N(\lambda) =\braket{\lambda|\lambda}= \bar F[\bar \Lambda]*F[\Lambda],
\end{equation}
where $\bar \Lambda$ is the complex conjugate matrix to the $\Lambda$ diagonal matrix, we find after a straightforward computation that:
\begin{equation}
\bra{\lambda} \hat H \ket \lambda = \sum \lambda^i \partial_{\lambda^i} \log(N(\lambda)) = \sum_i \lambda^i p_i,
\end{equation}
where $p_i$ are a new set of variables, given by derivatives of $K=\log(N(\lambda))$. 

Similarly, the Berry phase term is given by:
\begin{eqnarray}
\lim_{\tilde \lambda\to \lambda} \frac{\bra {\tilde \lambda}}{\sqrt{N(\tilde \lambda)} } i \partial_t  \frac{\ket {\lambda}}{\sqrt{N(\lambda)} }&=& \lim_{\tilde \lambda\to \lambda}  \braket{\tilde \lambda| \lambda}\frac 1{\sqrt{N(\tilde \lambda)}} i\partial_t \frac1{\sqrt{N(\lambda)} } + \frac{1}{\sqrt{N(\tilde \lambda) N(\lambda) }}\braket{\tilde \lambda | i \partial_t| \lambda}\nonumber\\
&=& i\sum_i \dot \lambda^i \partial_i \log(N(\lambda)) +\hbox{total time derivative} \nonumber \\
&\simeq& \sum i p_i \dot \lambda^i, 
\end{eqnarray}
where we drop the term that is a total derivative of $\log(N)$, since total derivatives do not contribute to the action.
The action we get is surprisingly simple:
\begin{equation}
S= \int dt \sum i p_i \dot \lambda^i - p_i \lambda_i. 
\end{equation}

In this equation, we should think of $p_i$ as the canonical conjugates of $\lambda^i$.
Solving the equations of motion immediately gives $\lambda_i = \lambda_i(0) \exp(-i t)$, which is the correct classical behavior for the coherent state fields. 
The $\lambda^i$ are well defined for the HCIZ integral, but when we consider some of the generalizations, we realize that we should use the $A^i$ variables rather than the $\lambda$ directly. The $A^i$ variables are products of $\Lambda$ for a quiver; only the $A$ eigenvalues enter $N$. We want the $p_i$ variables to be independent variables, so we rewrite the action in terms of the $A$ variables:
\begin{equation}
S= \int dt\sum i p_i \dot A^i -  n p_i A^i, 
\end{equation}
where $n$ is the number of nodes in the quiver diagram.

A more pressing question is evaluating $p_i$. We turn to \eqref{eq:CHIZsaddle}, which allows us to compute $p_i$. We want to express the result in the saddle point approximation and check where it is valid.
Consider a single large eigenvalue parameter of $\lambda^1$ (and make the others as small as needed). We want to know which saddles contribute. It is clear that the exponential in the saddle satisfies the following inequality:
\begin{equation}
|\exp( \bar \lambda^{\sigma(1)} \lambda^1+\dots) |< |\exp( \bar \lambda^1 \lambda^1 +\bar \lambda^2 \lambda^2+\dots) |,
\end{equation}
which can be proved by the Cauchy-Schwearz inequality.
We see that we require $\pi(1)=1$ . If we add more large eigenvalues, it becomes obvious that the dominant saddle is the one of the identity permutations by the same method. 
Keeping in mind that the other $\lambda$ are small, our expression for the overlap is approximated by:
\begin{equation}
\log(N(\lambda^1, \bar \lambda^1) ) \sim  \lambda^1\bar \lambda^1 -(N-1) \log( \bar \lambda^1\lambda^1),
\end{equation}
where we only take into account the $\lambda^1$ dependence and the dominant saddle.
We find this way that:
\begin{equation}
p_1 \lambda^1 = \bar \lambda^1 \lambda^1 -( N-1)
\end{equation}
We expect the energy of the configuration to be positive. This gives a lower bound on the collective coordinate $\lambda$, as follows:
\begin{equation}
 \bar \lambda^1 \lambda^1> N,
\end{equation}
where we have taken the large $N$ approximation. This is the complementary regime to the one found in \eqref{eq:range}. The analysis in \cite{Berenstein:2014zxa} looked directly at the expansions of truncated exponentials to do this. Here, we see that the saddle point encodes this 
information systematically. More importantly, if we say that $\lambda \sim \sqrt N$ in a scaling sense, then the energy stored in the state is of order $N$. 
This is usually associated with the energy of a D-brane. Indeed, these eigenvalues must be collective coordinates for AdS giant gravitons.

We also notice that $\bar \lambda \lambda$ is the Kahler potential of a flat manifold. This is recovered easily in the present formulation.
Let us define:
\begin{equation}
K= \log( N(\bar \Lambda, \Lambda))
\end{equation}

The wave functions we constructed begin as holomorphic functions of the $\lambda^i$. 
Therefore, they already represent a complex structure of the parameters. The action we wrote induces a symplectic structure on the $\lambda, p$ coordinates; we note that $p$ is essentially $\bar \lambda$.
We find that the symplectic form (expressed in terms of the $\bar z, z$ coordinates) is given by:
\begin{eqnarray}
\omega= d( \sum p_i d z^i) &=& d \sum \partial_i K d z^i= d( \partial K) = (\partial+\bar \partial) \partial K \nonumber\\
&=& (\bar \partial_j\partial_i K ) d \bar z^j \wedge dz^i = \sum_i  d \bar z^i \wedge dz^i
\end{eqnarray}
We find from the dominant saddle point computation that if we include all eigenvalues, the logarithmic correction to $K$ does not affect the metric. The metric is flat. This matches what we expect. The eigenvalues $\lambda$ parametrize the coherent states we have described. More importantly, in the semi-classical limit, the coherent states encode a particular field theory configuration, where the field we are quantizing ($Z$ in this case) has a classical vacuum expectation value $Z\sim \Lambda$. 
Notice that because the HCIZ integral is exactly a sum over saddles, all corrections to $K$ that could further arise are the contribution of other saddles: they are to be considered as a non-perturbative effect. 

A similar result holds for the determinant calculation, where we would again find the flat K\"aler metric, as the overlap is a simple exponential.
The one difference is how to compute the energy. When we count powers of $(a^\dagger)^k$ in the determinant, they are paired with powers of $\lambda^{N-k}$. In this case, the Hamiltonian is therefore $N-\lambda\partial_\lambda$. Putting everything together, we find that the energy in this case is:
\begin{equation}
H_{det}= N- \bar\lambda \lambda= \bar \phi \phi.
\end{equation}
For the energy to be greater than zero, we need $\bar \lambda \lambda <N$, which coincides with the condition that the Hubbard-Stratonovich field $\phi$ has a good saddle. Again, if we look at the scaling of the energy, it scales like $N$. This again must be interpreted as a D-brane. These are the sphere giant gravitons.

Let us now turn to the $A_{n-1}$ quivers. We need to understand the asymptotic expansion of the generalized hypergeometric function (see also \cite{Zinn-Justin:2002rai}) for the dominant saddle:
\begin{equation}
\Phi_n(A \bar A) = \sum_m  \frac{1}{(m!)^n}(A \bar A)^m,
\end{equation}
which replaces the exponential; in the denominator, we have $(A\bar A)^N$. 
Notice that since this only depends on $A= \prod_{i=1}^n (\lambda_{(i,i+1)} ^1)$, we can take all $\lambda$ variables to be identical (we should do this anyhow as they are not gauge invariant on their own). We can do the same with $\bar \lambda$.
From our integral representation of $\Phi$, we find that:
\begin{equation}
\Phi_n(A \bar A) = \frac{1}{(2\pi)^n}\iint \prod d\theta_i \exp( \lambda \bar \lambda \exp(i (\theta_i-\theta_{i+1})) )
\end{equation}
Now, $\bar\lambda\lambda$ should be large, so we can saddle the integral over the angles. The critical point for the maximum is when all the angles are equal to one other. At this point, the saddle is for real values of $\theta$. If the $\lambda$ are not equal to one other, then the saddles move in the complex plane, and we will need to worry about both the real and imaginary parts of the variables $\theta$.

Asymptotically, we find that:
\begin{equation}
\Phi_n(A \bar A) \sim \exp( n \lambda \bar \lambda) = \exp( n (A\bar A)^{1/n})
\end{equation}
We obtain two results from this. Our first result is:
\begin{equation}
p \sim \bar A (A\bar A)^{1/n-1}  -  N /A,
\end{equation}
where we have included the measure term. In this case, $p$ is not the complex conjugate of $A$ and a holomorphic correction. The energy function is $\sum \lambda_{i,i+1} \partial_{i,i+1}$. This measures the degree with respect to $\lambda$. Since $A$ is composed of a product of $n$ lambdas, we need to multiply the degree of $A$ by $n$ to get the Hamiltonian.
We find that the effective action is
\begin{equation}
S= i \int p \dot A - (n) p A= \int dt \left(i  p \dot A - n [(\bar A A)^{1/n} - (N-1)]\right)
\end{equation}
as expected. The K\"ahler potential gives rise to a flat geometry, but the complex structure is that of ${\mathbb C}/{\mathbb Z}_n$, as one would expect from the quiver theory. We get a similar constraint from the positive energy of a single large eigenvalue, namely that $\bar \lambda \lambda>N-1$. 
The K\"ahler form we get for the one eigenvalue is:
\begin{equation}
\omega = \frac1 n (\bar A A)^{1/n-1} d\bar A \wedge d A
\end{equation}
Notice that these coherent states also match the generalized coherent states introduced in \cite{Berenstein:2015ooa} that were constructed combinatorially.
That the energy is essentially the K\"ahler potential is a semiclassical result expected for BPS states \cite{Berenstein:2005aa,Berenstein:2007wi}

Consider now a $U(1)\times U(N+1)$ theory in the example where we showed that naive localization does not work. We still get a function similar to $\phi$, but now we have:
\begin{equation}
\Phi= \sum_{m=0}^\infty \frac{1}{m!(m+N)!} (A \bar A)^m = (A\bar A)^{-N/2} I_N(2 \sqrt{A\bar A})
\end{equation}
and because one of the terms is a $U(1)$, there is no denominator. The hypergeometric function, when written in terms of Bessel functions, looks as if it does have a denominator, with $N/2$ other eigenvalues. It is as if gauging $U(N)$ only counts for half as many eigenvalues. 
We want to understand the corresponding expressions at medium to moderate large $N$. 

There are two regimes we want to consider: large $A$ and small $A$. For large $A$, we need to use the asymptotic expansion of the Bessel function. 
This time, we choose to start from a different integral representation of $I_v(z)$ and use Watson's lemma to compute the asymptotic expansion of $I_v(z)$. We begin with:
\begin{equation}
    I_v(z) = \frac{(2z)^{-1/2} e^z}{\sqrt{\pi}\Gamma(v+\frac{1}{2})}\int_0^1e^{-2zt}t^{v-\frac{1}{2}}(1-t)^{v-\frac{1}{2}}dt,
\end{equation}
where we've stipulated that $\mathcal{R}e( v )> -\frac{1}{2}$. We use Watson's lemma, which holds that an integral $F_\lambda(z)$, defined such that:
\begin{equation}
    F_\lambda(z) = \int_0^\infty t^{\lambda-1}f(t)e^{-zt}dt,
\end{equation}
where $\mathcal{R}e (\lambda) > 0$ and the function $f(t)$ has a  Taylor series expansion :
\begin{equation}
f(t)\approx \sum_{n=0}^\infty a_nt^n,
\end{equation}
as $t\rightarrow 0_+$, can be approximated as:
\begin{equation}
F_\lambda(z) \approx \sum_{n=0}^\infty a_n\frac{\Gamma(n+\lambda)}{z^{n+\lambda}}
\end{equation}
in the large argument limit for $z$. 
For our case, in the large argument limit, our asymptotic expansion of $I_v(z)$ becomes:
\begin{equation}
    I_v(z) = \frac{e^z}{\sqrt{2\pi z}\Gamma\left(v+\frac{1}{2}\right)}\sum_{k=0}^\infty (-1)^k{{v-\frac{1}{2}}\choose{k}}\frac{\Gamma\left(k+v+\frac{1}{2}\right)}{z^k}\label{eq:besselexp}
\end{equation}
Putting it all together, we get that:
\begin{equation}
\log(\phi) \sim  2(\bar A A)^{1/2} -(N/2+1/2) \log(A\bar A) -\frac{4N^2-1}{2\sqrt{\bar AA}}, 
\end{equation}
so for very large $A$, we once again arrive at the K\"ahler potential of ${\mathbb C}/{\mathbb Z}_2$ when we are allowed to ignore the $1/\sqrt{A\bar A}$ term.
The expression above is valid when the second term is smaller than the first in the series \eqref{eq:besselexp}, or when $(4N^2-1)/(2\sqrt{\bar AA})^{-1} <1$. 
So $A\bar A \sim N^2$ has a $N^2$ scaling. This is equivalent to $\sqrt A\sim \sqrt N$, which gives the estimate in terms of the naive eigenvalues $\lambda$, rather than the composite object $A$.

Notice, however, that there is a large correction at large $N$.
For small $A$, we use the series directly to find that:
\begin{equation}
\Phi \sim \sum_{n} \frac{1}{n! N^n} (A\bar A)^n \sim \exp( A\bar A/N),
\end{equation}
where we have used a large $N$ approximation for the denominator. This way, we have:
\begin{equation}
\log (\Phi) \sim (A\bar A)/N,
\end{equation}
so the metric becomes that of the flat complex plane, rather than a cone. This applies so long as $A\bar A$ is not too large, so that the approximation of the denominators is valid near the terms of the series that contribute the most.
This means that the tip of the cone is flattened (rounded) over a rather large range. We can think of this geometric calculation as quantum effects deforming the singularity away in a manner analogous to \cite{Klebanov:2000hb,Gopakumar:1998ki}, where the extra ``fractional" branes at the tip of a cone lead to a measurable deformation of the geometry.
 
This effect can be thought of as a toy model of geometric transitions. The crossover from the small $A$ to the large $A$ happens when the formulas are roughly comparable to each other; that is, when $(\bar A A)^{1/2} \sim A\bar A/N$, or equivalently, where $(A\bar A)^{1/2}\sim N$, which is the same estimate we obtained from the asymptotic series.

\subsection{More open strings}

In our discussion so far, we have been able to introduce open strings for the sphere giant graviton D-brane (the determinant). We now want to do this for the coherent state excitation parameterized by a few large eigenvalues $\lambda^i$. We notice that in the space of matrices associated with $\Lambda$, we already have a diagonal matrix; to each eigenvalue, we can associate an eigenvector ${\ket v}_i$. When we multiply the matrix by $U \Lambda U^{-1}$, this vector is rotated to $U \ket v_i$. We can stretch a sting between eigenvalues $i j$ if we add words $W$ as follows:
\begin{equation}
\ket{\Lambda, W}\sim \int dU \exp\left(\tr({U \Lambda U^{-1}} a^\dagger)\right)  \bra v_j U^{-1} W U  \ket v_i \ket 0,
\end{equation}
We may add more words in a similar manner.
Notice that we are still doing integrals over $U(N)$ and that now the vectors $v$ transform with the phases of $U(1)^N$ that do not belong to the coset space $U(N)/U(1)^N$.
The total phase must cancel, so we find that for each  $\ket v_i$, there must be another $\bar v_i$ somewhere else, which is to say that we get the Gauss' law constraint for $U(1)^N$. This is as expected. The Coulomb branch in $Z$ breaks the gauge group to $U(1)^N$, tied to diagonalizing the expectation values of $Z$. When we consider saddles in the integral, the saddles in $U$ are very pronounced if for a single eigenvalue $|\lambda \bar \lambda| $ that is large of order $N$. This means that we can ignore contributions from the $W$ to the saddles. Instead, we evaluate $U$ in the corresponding permutation matrix and keep the integration over the phases explicit.
This procedure can lead to an effective action of D-branes with strings. Let us understand this for a pair of large eigenvalues $\lambda_{12}$ and the word $a_X^\dagger$. We need to add the contribution $\bra v_1 U^{-1} a_X^\dagger U \ket v_2$. Now we act with the 1 loop effective Hamiltonian in ${\cal N}=4 $ SYM, which is written as a contribution coming strictly from F-terms as in \cite{Berenstein:2002jq}  (here we use a slightly modified version of \cite{Beisert:2002ff} with raising and lowering operators, similar to the notation in \cite{Berenstein:2004ys}):
\begin{equation}
g_{YM}^2 \tr \left([a_X^\dagger,a^\dagger_Z][a_Z,a_X]\right)
\end{equation}
When acting on the simple word, the extra lowering operator $a_Z$ brings down a copy of $U \Lambda U^{-1}$, either to the left or to the right.
For more general states, the reader may refer to the works \cite{Minahan:2002ve,Beisert:2003jj} (see \cite{Beisert:2010jr} for a review of the integrability program).
These two pieces come as: 
\begin{equation}
g_{YM}^2\bra v_1 U^{-1} U \Lambda U^{-1} [a_X^\dagger,a^\dagger_Z] U\ket v_2-\bra v_1 U^{-1}  [a_X^\dagger,a^\dagger_Z]  U \Lambda U^{-1}\ket v_2
\end{equation} 
Cancelling the $U$ and noting that $\ket v_{12}$ are eigenstates of $\Lambda$, we obtain the answer:
\begin{equation}
g_{YM}^2(\lambda_1-\lambda_2) \bra v_1 U^{-1} [a_X^\dagger,a^\dagger_Z] U\ket v_2
\end{equation} 
We now perform the same trick when computing with the dual vector, so that $a_Z^\dagger$ brings down powers of $\bar \Lambda$. Again, we obtain an integral that involves $U^{*-1}$, and we pick the identity saddle, so that we get:
\begin{equation}
H_{1-loop} \sim g_{YM}^2| (\lambda_1-\lambda_2)|^2\label{eq:openen}
\end{equation}
With the $N$ scaling of $\lambda$, we get a finite contribution in the t'Hooft limit. This is exactly what is expected from other approaches that are based on coherent states without the $U$ integral formalism \cite{Berenstein:2013eya, Berenstein:2014zxa, Berenstein:2020grg,Berenstein:2020jen}. This should be contrasted with the difficult combinatoric computations that lead to ``open spring theory" \cite{deMelloKoch:2011ci}, which contains the same physics.
More generally, such correlators of matrix elements of $U$ have been studied  in \cite{Shatashvili:1992fw,Morozov:1992zb,Eynard:2005wg}, where some exact expressions can be found. It should be interesting to develop that further.

Developing this idea further is beyond the scope of the present paper.
The eventual goal of such a program would be to simplify the types of analysis found in the works \cite{deMelloKoch:2020agz,deMelloKoch:2020jmf}.
It is likely that this basis is close to the so called Gauss graph basis described in \cite{deMelloKoch:2012ck,deMelloKoch:2017ytj}. 

There is a second problem we would like to discuss: how to include more than one sphere giant graviton in this discussion. 
The idea of how to do it correctly should be motivated by what we have seen already, namely, that there is a good character expansion formalism.
For sphere giants, we should have a generalization of the type:
\begin{equation}
G[M] \sim \sum_R \frac 1{s_R} \chi_R(a^\dagger_z) \chi_{R^T} (M^{-1}),
\end{equation} 
where the characters of the generalized eigenvalues $M$ involve the dual partition $R^T$. We are also including some denominator expressions $s_R$ in case they are needed.
We should also have inverse powers of $M$; as in our example of a single sphere giant, counting $a^\dagger$ runs opposite to counting powers of $\lambda$. 
Alternatively, we can go ahead and introduce new variables $u= 1/\lambda$; then, counting powers of $u$ counts the powers of $a^\dagger$. Curiously, inversion of coordinates also seems to play an important role for the spin chain in the $SL(2)$ sector for open strings \cite{Berenstein:2020jen}. Based also on the introduction of fermions to deal with a single determinant, we would expect that such a generalization involves a fermion integral with more fermions. The proportionality constant in front of $G$ should be $\lambda^N$ for a single eigenvalue (that is, when $M$ is of rank 1).
This suggests using $(\det M)^N$ as a normalization factor more generally, which is suggestive of an integral over fermionic rectangular matrices of size $\text{rank}(M)\times N$.

In the paper \cite{Berenstein:2013md}, multiple giants are introduced as products of determinants. We will justify this idea further. There is an 
expansion in characters from an algebraic identity (see for example \cite{Morozov:2009jv}): 
\begin{equation}
\sum_R \chi_R(t)\chi_R(a_Z^\dagger) = \frac 1{\det\left(I\otimes I -t \otimes a_Z^\dagger \right)}
\end{equation}
where $t$ is an arbitrary $M\times M$ matrix that is tied to the ``coordinates" of M AdS giant gravitons. A straightforward evaluation of the norm of the left shows that the state is not normalizable unless $t=0$. This is because the operators $a_Z^\dagger$ are unbounded, so the Taylor expansion around $t=0$ is not convergent.
In this sense, this is a formal expansion.
 
Because the formula is an inverse determinant, it can be written as a bosonic integral. In the study of Wilson loops, it is noted that if one bosonizes a determinant Wilson loop, one passes from single column Young tableaux to row Young tableaux \cite{Gomis:2006im}, thus switching between two types of ``dual" D-brane representation: D5 and D3 branes. Here we see how the same idea can be written; the idea is then to fermionize the determinant above written as a bosonic integral to find:
\begin{equation}
\det\left(I\otimes I -t \otimes a_z^\dagger \right)= \int d\bar \chi d \chi \exp\left( \tr(\bar \chi \chi - \bar \chi a_Z^\dagger \chi t ) \right)
\end{equation}
In this case, to each eigenvalue $t_s$, we associate a flavor of fermions. For each such $t_s$ eigenvalue, we can isolate fermions on the left and and on the right, and thus get the open strings attached to different giants. This would reproduce the ideas of \cite{Berenstein:2013md}, but would use the fermion language. To generalize the calculation, when one considers the Hubbard-Stratonovich trick, one should introduce an $M\times M$ collective field. One would like to have a saddle of this field that aligns in the ``identity" permutation. The $t$ variables here are exactly like the inverse of the $\lambda$ parameters.  Notice also that in matrix quantum models of rectangular matrices with fermions, one usually ends up pairing representations with dual young diagrams of the two groups under which the matrices transform \cite{Berenstein:2004hw,deMelloKoch:2012sie,Berenstein:2019esh}.
 
We can also check that if we use the second Cauchy identity for Schur functions, we arrive at:
\begin{equation}
\sum _R \chi_R(t) \chi_{R^T} (a_Z^\dagger)= \det\left(1+t\otimes a_Z^\dagger \right)
\end{equation}
so that up to a sign, we get the correct generating series that we wanted in terms of characters. 

Now, adding open strings becomes very simple. In the basis where $t$ is diagonal, we can have fermions $\psi^i$ and $\bar \psi_j$. Sandwiching words between these fermions allows one to form general states with open strings. The Gauss' law constraint becomes trivial: there is a $U(1)^M$ charge under which the fermion integral is invariant. This symmetry sends $\bar \psi_j \to \exp(-i \theta_j)\bar \theta_j$ and $\psi^j \to \exp(i \theta_j) \psi^j$, one for each parameter  $t_j$. We find that there needs to be as many $\psi^j$ as $\bar \psi_j$, that is, the same number of positively charged and negatively charged particles with respect to the D-brane $U(1)$ charge. When two of the $t$  coincide, there is an enhanced $U(2)$ symmetry of the fermionic integral. 
This should be the generating series counterpart of how to attach strings to sphere giants combinatorially by adding boxes to Young diagrams  \cite{Balasubramanian:2004nb,deMelloKoch:2007rqf}. 

There should also be a more general theory of coherent states that has both sphere giants and AdS giants appearing more democratically, as one expects from the strict infinite N limit, where they can be constructed directly by making note of the symmetry between Young diagrams and their transposes \cite{Berenstein:2017abm}.
In the strict $N\to \infty$ limit this can be made very precise. For example, the overlaps of multiple giants   \cite{Lin:2017dnz} are easily seen to be given by formulae that can be expanded in terms of Cauchy's character formulae. The parameters of those coherent states play the role of eigenvalue collective coordinates.
A general setup for finite $N$ that is democratic between these probably involves both fermion and boson integrals. It is likely that there is a supermatrix model that does this.

\section{Coherent states  for $1/4$ and $1/8$ BPS states}\label{sec:QBP}

We are now ready to tackle coherent states for $1/4$ and $1/8$ BPS bosonic states. The idea, following our previous development, is to start with averaged coherent states:
\begin{equation}
F[\Lambda_Z, \Lambda_X,\Lambda_Y] =  \frac{1}{Vol} \int dU \exp\left( \tr( U\Lambda_X U^{-1} a_X^\dagger+U\Lambda_Y U^{-1} a_Y^\dagger+U\Lambda_Z U^{-1} a_Z^\dagger )\right)\ket 0
\end{equation}
We now want to concentrate on the states that are $1/4$ and $1/8$ BPS at one loop order. The effective Hamiltonian is given by:
\begin{equation}
H=\tr\left([a_X^\dagger,a^\dagger_Y][a_Y,a_X]\right)+\hbox{cyclic},
\end{equation}
and since $H$ is a sum of squares, we get $H\geq 0$ as an operator. When we let $H$ act on $F$, we see that we get a result that is identically equal to zero when the $\Lambda$ matrices commute. 
We thus insist that the parameters $\Lambda_X, \Lambda_Y, \lambda_Z$ are commuting matrices, as they should be. The coherent states are a semiclassical approximation to expectation values of fields. The moduli space of vacua occurs when the classical expectation values of the fields (after gauge fixing) commute. 
Namely, we need the expectation values to commute: $[X,Y]=0 \dots$. 

Now we perform our usual manipulations contracting $a_X^\dagger, a_X$ etc, to find the overlap:
\begin{eqnarray}
\bar F[\bar\Lambda_Z, \bar\Lambda_X,\bar \Lambda_Y]*F[\Lambda_Z, \Lambda_X,\Lambda_Y] &=&\\
  \frac{1}{Vol} \int dU \exp\left( \tr( U\Lambda_X U^{-1} \bar \Lambda_X+U\Lambda_Y U^{-1}\bar \Lambda_Y+U\Lambda_Z U^{-1}\bar \Lambda_Z)\right)\ket 0
\end{eqnarray}
This answer has a manifest $U(3)$ symmetry of rotations where the matrices $\Lambda_X,\Lambda_Y,\Lambda_Z$ transform as a $3$ of $U(3)$, and the conjugate $\bar \Lambda$ transform as the $\bar 3$.

Here are a couple of observations. First, the action in the integral is also evaluated on a $U(N)/U(1)^N$ geometry; thus, the phases of $U$ acting on the left disappear. 
Some of the critical points of $U$ are the same as those in the HCIZ integral: permutation matrices.  
There may be additional ones. The conditions for critical points are:
\begin{equation}
[\bar \Lambda_X, U\Lambda_X U^{-1}]+  [\bar \Lambda_Y, U\Lambda_Y U^{-1}]+[\bar \Lambda_Z, U\Lambda_Z U^{-1}]=0
\end{equation}
It is clear that when $U$ is a permutation matrix, they are critical points.

If $\bar \Lambda$ and $\Lambda$ are real, then the term in the exponential is real. In that case, Morse theory for the compact manifold over which we are doing the integral suggests that for small enough perturbations in $\Lambda_X,\Lambda_Y$ slightly away from zero, the set of isolated critical points does not change (these depend continuously on the action when thought of as a Morse function on the manifold we are interested in).
That means that we can evaluate a 1-loop approximation around the same saddles and get an approximation for the overlap. These should be dominant.
If the integral above is localizable (which we have not proved), then the sum over all saddles (including complex saddles) is exact. Considering that one can localize on $1/8$ BPS Wilson loops in \cite{Pestun:2009nn}, the idea that an integral like the one above or a variation of it is amenable to exact localization is very plausible.
If additional saddles are needed, they will be complex. 

We will do the naive saddle sum now, over the saddles we know. We find that for each saddle, we have a permutation matrix $U\sim P$, and the saddle gives:
\begin{equation}
S_{\pi}= \sum_i \lambda_X^i \bar \lambda_X^{\pi(i)} +\lambda_Y^i \bar \lambda_Y^{\pi(i)} + \lambda_Z^i \bar\lambda_Z^{\pi(i)}.
\end{equation}
The square root of the measure at the saddle is evaluated readily to a product:
\begin{equation}
\mu_\pi = \prod_{i<j} (\vec \lambda_i -\vec \lambda_j)\cdot(\vec {\bar \lambda}_{\pi(i)} -\vec {\bar\lambda}_{\pi(j)}) 
\end{equation}
Such measures appear in the computation of the volume of the gauge orbit in \cite{Berenstein:2005aa} (they can also be extended to other orbifolds or more general setups \cite{Berenstein:2006yy,Berenstein:2007wi,Berenstein:2007kq}). Unlike in the case of the HCIZ integral, this measure does not factorize holomorphically. Moreover, different saddles have different denominators. This suggests that some poles will not cancel to give rise to a polynomial in the $\lambda$ variables.
If the poles are not cancelled, then there are two possibilities: either localization does not work, or there exist additional complex saddles in the complexified $U$ variables that need to be taken into account.

The measure does reduce to the product of Vandermonde determinants when we get rid of the $X,Y$ variables $\lambda_X=\lambda_y=0$. It is also invariant under the $SU(3)$ rotations in $X,Y,Z$, so the answer is consistent from the $SU(3)$ group theory considerations.
Notice that $\mu^2$ does change sign with the permutations in the same way the Vandermonde does, so the accompanying sign in the saddle point evaluation at $\lambda_x=\lambda_y=0$ should be kept. We find an approximation given by
\begin{equation}
\bar F[\bar\Lambda_Z, \bar\Lambda_X,\bar \Lambda_Y]*F[\Lambda_Z, \Lambda_X,\Lambda_Y]= \sum_\pi (-1)^\pi\frac{\exp(S_\pi)}{\mu_\pi} 
\end{equation}
We can also try to go to our collective coordinate formulation. In that case, $\bar \Lambda$ is the adjoint of $\Lambda$, and the trivial permutation dominates (which one can show by the Cauchy-Schwarz inequality).
In that limit, we find that:
\begin{equation}
\bar F[\bar\Lambda_Z, \bar\Lambda_X,\bar \Lambda_Y]*F[\Lambda_Z, \Lambda_X,\Lambda_Y] \sim \frac{\exp(\tr( \bar \Lambda \Lambda)) }{\prod_{i<j} |\vec \lambda_i -\vec \lambda_j|^2},
\end{equation}
and that the energy would evaluate to:
\begin{equation}
\tr( \bar \Lambda \Lambda) - \lambda \partial_\lambda \ln({\prod_{i<j} |\vec \lambda_i -\vec \lambda_j|^2}) = \tr( \bar \Lambda \Lambda) - N(N-1),
\end{equation}
where we are using the fact that the measure is a homogeneous function (see \cite{Berenstein:2005aa}). For a single large eigenvalue to be well behaved, we require that the energy in that eigenvalue (evaluated with $N-1$ eigenvalues set to zero) is positive compared to the result when $N\rightarrow N-1$.
This gives a $SU(3)$ covariant version of the eigenvalue being larger than $N$. That is:
\begin{equation}
\vec {\bar \lambda}\cdot \vec \lambda >N-1
\end{equation}
We know that the equation of motion of $\lambda_i$ is $i\dot \lambda_i=\lambda_i$, and that the Hamiltonian is essentially $\sum \bar \lambda_i \lambda_i$ up to a constant. 
This suggests that the canonical conjugate to $\lambda_i$ is $\bar \lambda_i$. Indeed, an action based on that prescription alone would give the correct equations of motion. Moreover, the K\"ahler potential would be that of a flat geometry. Notice, however, that we have a quantum correction from the measure.

This is important. One way of interpreting this correction has to do with the counting of states. When we write the generating function for a single large eigenvalue, we get:
\begin{equation}
\bar F*F \sim \exp( \bar \lambda \lambda)/(\bar \lambda \lambda)^2\rightarrow \sum_{m=0}^\infty N!\frac{(\bar \lambda \lambda)^m}{(m+N)!},
\end{equation}
where we only keep the regular part of the answer, as the full generating series has no singularities at the origin. In this notation, we are suppressing the $SU(3)$ labels, and we only keep the polynomial part of the answer. 
We see that there is one state per monomial $\lambda^{n_1}_X \lambda^{n_2}_Y\lambda^{n_3}_Z$ from expanding each of the terms of the sum; the expansion gives a sum of squares. The number of states at energy $k$ is the number of states of the k-th completely symmetric representation of $SU(3)$. This has dimension $(1+k)(2+k)/2$. Semiclassically, this should be the volume of phase space between energy $k$ and $k+1$; our phase space is the complex manifold we are discussing.
We should now substitute $k=E\sim \bar \lambda \lambda-N=r^2-N$, where $r^2$ is the norm. We find that the volume at fixed $r$ scales like $(r^2-N)^2 r dr/2$, which is different from that of a flat geometry at the origin. At large $r$, the correction does not matter and the metric becomes scaling, but at finite $r$, we get the wrong counting of states. The representation theory has also been studied directly in \cite{Biswas:2006tj}, using different methods.

If we have a few branes with large eigenvalues, the measure gets a correction from the product of two measures that is singular when the large eigenvalues coincide. This indicates the enhanced symmetry of the integral-- when the eigenvalues coincide, the critical point leaves $SU(2)$ invariant, and the fixed point is not isolated. This is associated with the enhanced gauge symmetry of coinciding branes.

The main point we are making is that the idea of saddles dominating is still accurate and the type of constructions that are used for attaching strings to these setups still hold, including energies like equation \eqref{eq:openen}, properly covariantized to be $U(3)$ invariant. The BPS states here are only implicit. They are expressed as polynomials of the eigenvalues $\lambda$ multiplying functions of oscillators that we have not computed explicitly, and written as integrals over the group. 
They are also invariant under combined permutations of the eigenvalues. This should be contrasted with other approaches to this problem \cite{Pasukonis:2010rv,Lewis-Brown:2020nmg}.
 
Another option is to keep $\Lambda_Z$ finite and expand the exponential in the other variables $\Lambda_{X,Y}$ as a power series. Results to each order would then be given in terms of correlators of $U,U^{-1}$  matrix elements in the ``HCIZ ensemble." Such correlators have been studied in \cite{Shatashvili:1992fw,Morozov:1992zb,Eynard:2005wg} and they do correspond to elaborate sums over the saddles. Such formal expansions would add open strings to the eigenvalues of  the $\Lambda_Z$ configuration and would tie in with the open string formalism briefly mentioned in this paper.

\section{Discussion}\label{sec:DIS}

In this paper, we have discussed a new application of the Harish-Chandra-Itzykson-Zuber integrals and their generalizations to study correlators of BPS states in ${\cal N}=4$ SYM. 
The main idea was to introduce coherent states and average them over a group orbit to obtain gauge invariant states. We were able to reproduce various results that were obtained originally from combinatorial arguments.
We then promoted the parameters in the generating function to collective coordinates of states. We exploited the fact that the integrals in question are written as sums of saddles to find that there is a dominant saddle.  
This allows one to make various approximations to find the effective action of these collective coordinates directly. We showed how a similar structure could be found with determinants that also lead to dominant saddles and explained why this is the correct generalization in terms of the type of algebraic structure that arises.

We also showed how to introduce open strings in all of these these setups. 
We found that computations done this way require knowledge of correlators of matrix elements of unitary matrices in the HCIZ ensemble. This method provides a complementary approach to study anomalous dimensions of open strings that we are currently investigating.
Again, the fact that there is a dominant saddle for the setup allows simplifications to be made in computations. We demonstrated that these open strings need to satisfy a Gauss' law constraint that becomes evident in the formulation, including the non-abelian enhancement that occurs when D-branes coincide.

We extended these ideas to more general group integrals for $1/4$ and $1/8$ BPS states. We believe that because $1/8$ BPS WIlson loops are computable using localization, one can find a formalism where the correlators of these states are computed this way, rather than the combinatorial approaches found in current literature.

It would be interesting if similar ideas can be used to study the ABJM theory \cite{Aharony:2008ug} correlators. In that case, Wilson loops can be computed with supermatrix models \cite{Drukker:2009hy}, derived using localization methods \cite{Drukker:2010nc}. The associated spin chain is integrable \cite{Minahan:2008hf}.

We have not yet applied our ideas to the study of higher point functions along the lines of \cite{Beem:2013sza} which may also be computed using localization methods, nor to the fact that the HCIZ integrals are also tau functions of integrable systems (see \cite{Morozov:2009jv}). All of these avenues suggest a rich setup of possible applications of the ideas presented in this paper to the computation of protected quantities in various setups.

\acknowledgments
We would like to thank A. Holguin for discussions. D.B. would also like to thank S. Komatsu for discussions. 
D.B. would also like to thank the KITP workshop on Confinement, Flux Tubes, and Large N. Work supported in part by the Department of Energy under grant DE-SC 0011702.
This research was supported in part by the National Science Foundation under Grant No. NSF PHY-1748958.

\bibliographystyle{jhep}

\bibliography{coherent}

\providecommand{\href}[2]{#2}\begingroup\raggedright\begin{thebibliography}{10}

\bibitem{Corley:2001zk}
S.~Corley, A.~Jevicki and S.~Ramgoolam, \emph{{Exact correlators of giant
  gravitons from dual N=4 SYM theory}},
  \href{https://doi.org/10.4310/ATMP.2001.v5.n4.a6}{\emph{Adv. Theor. Math.
  Phys.} {\bfseries 5} (2002) 809}
  [\href{https://arxiv.org/abs/hep-th/0111222}{{\ttfamily hep-th/0111222}}].

\bibitem{Lee:1998bxa}
S.~Lee, S.~Minwalla, M.~Rangamani and N.~Seiberg, \emph{{Three point functions
  of chiral operators in D = 4, N=4 SYM at large N}},
  \href{https://doi.org/10.4310/ATMP.1998.v2.n4.a1}{\emph{Adv. Theor. Math.
  Phys.} {\bfseries 2} (1998) 697}
  [\href{https://arxiv.org/abs/hep-th/9806074}{{\ttfamily hep-th/9806074}}].

\bibitem{Baggio:2012rr}
M.~Baggio, J.~de~Boer and K.~Papadodimas, \emph{{A non-renormalization theorem
  for chiral primary 3-point functions}},
  \href{https://doi.org/10.1007/JHEP07(2012)137}{\emph{JHEP} {\bfseries 07}
  (2012) 137} [\href{https://arxiv.org/abs/1203.1036}{{\ttfamily 1203.1036}}].

\bibitem{Rastelli:2016nze}
L.~Rastelli and X.~Zhou, \emph{{Mellin amplitudes for $AdS_5\times S^5$}},
  \href{https://doi.org/10.1103/PhysRevLett.118.091602}{\emph{Phys. Rev. Lett.}
  {\bfseries 118} (2017) 091602}
  [\href{https://arxiv.org/abs/1608.06624}{{\ttfamily 1608.06624}}].

\bibitem{Dey:2011ea}
T.K.~Dey, \emph{{Exact Large $R$-charge Correlators in ABJM Theory}},
  \href{https://doi.org/10.1007/JHEP08(2011)066}{\emph{JHEP} {\bfseries 08}
  (2011) 066} [\href{https://arxiv.org/abs/1105.0218}{{\ttfamily 1105.0218}}].

\bibitem{Caputa:2012dg}
P.~Caputa and B.A.E.~Mohammed, \emph{{From Schurs to Giants in ABJ(M)}},
  \href{https://doi.org/10.1007/JHEP01(2013)055}{\emph{JHEP} {\bfseries 01}
  (2013) 055} [\href{https://arxiv.org/abs/1210.7705}{{\ttfamily 1210.7705}}].

\bibitem{Pasukonis:2013ts}
J.~Pasukonis and S.~Ramgoolam, \emph{{Quivers as Calculators: Counting,
  Correlators and Riemann Surfaces}},
  \href{https://doi.org/10.1007/JHEP04(2013)094}{\emph{JHEP} {\bfseries 04}
  (2013) 094} [\href{https://arxiv.org/abs/1301.1980}{{\ttfamily 1301.1980}}].

\bibitem{Berenstein:2015ooa}
D.~Berenstein, \emph{{Extremal chiral ring states in the AdS/CFT correspondence
  are described by free fermions for a generalized oscillator algebra}},
  \href{https://doi.org/10.1103/PhysRevD.92.046006}{\emph{Phys. Rev. D}
  {\bfseries 92} (2015) 046006}
  [\href{https://arxiv.org/abs/1504.05389}{{\ttfamily 1504.05389}}].

\bibitem{McGreevy:2000cw}
J.~McGreevy, L.~Susskind and N.~Toumbas, \emph{{Invasion of the giant gravitons
  from Anti-de Sitter space}},
  \href{https://doi.org/10.1088/1126-6708/2000/06/008}{\emph{JHEP} {\bfseries
  06} (2000) 008} [\href{https://arxiv.org/abs/hep-th/0003075}{{\ttfamily
  hep-th/0003075}}].

\bibitem{Balasubramanian:2001nh}
V.~Balasubramanian, M.~Berkooz, A.~Naqvi and M.J.~Strassler, \emph{{Giant
  gravitons in conformal field theory}},
  \href{https://doi.org/10.1088/1126-6708/2002/04/034}{\emph{JHEP} {\bfseries
  04} (2002) 034} [\href{https://arxiv.org/abs/hep-th/0107119}{{\ttfamily
  hep-th/0107119}}].

\bibitem{Grisaru:2000zn}
M.T.~Grisaru, R.C.~Myers and O.~Tafjord, \emph{{SUSY and goliath}},
  \href{https://doi.org/10.1088/1126-6708/2000/08/040}{\emph{JHEP} {\bfseries
  08} (2000) 040} [\href{https://arxiv.org/abs/hep-th/0008015}{{\ttfamily
  hep-th/0008015}}].

\bibitem{Hashimoto:2000zp}
A.~Hashimoto, S.~Hirano and N.~Itzhaki, \emph{{Large branes in AdS and their
  field theory dual}},
  \href{https://doi.org/10.1088/1126-6708/2000/08/051}{\emph{JHEP} {\bfseries
  08} (2000) 051} [\href{https://arxiv.org/abs/hep-th/0008016}{{\ttfamily
  hep-th/0008016}}].

\bibitem{Balasubramanian:2004nb}
V.~Balasubramanian, D.~Berenstein, B.~Feng and M.-x.~Huang, \emph{{D-branes in
  Yang-Mills theory and emergent gauge symmetry}},
  \href{https://doi.org/10.1088/1126-6708/2005/03/006}{\emph{JHEP} {\bfseries
  03} (2005) 006} [\href{https://arxiv.org/abs/hep-th/0411205}{{\ttfamily
  hep-th/0411205}}].

\bibitem{deMelloKoch:2007rqf}
R.~de~Mello~Koch, J.~Smolic and M.~Smolic, \emph{{Giant Gravitons - with
  Strings Attached (I)}},
  \href{https://doi.org/10.1088/1126-6708/2007/06/074}{\emph{JHEP} {\bfseries
  06} (2007) 074} [\href{https://arxiv.org/abs/hep-th/0701066}{{\ttfamily
  hep-th/0701066}}].

\bibitem{deMelloKoch:2007nbd}
R.~de~Mello~Koch, J.~Smolic and M.~Smolic, \emph{{Giant Gravitons - with
  Strings Attached (II)}},
  \href{https://doi.org/10.1088/1126-6708/2007/09/049}{\emph{JHEP} {\bfseries
  09} (2007) 049} [\href{https://arxiv.org/abs/hep-th/0701067}{{\ttfamily
  hep-th/0701067}}].

\bibitem{Bekker:2007ea}
D.~Bekker, R.~de~Mello~Koch and M.~Stephanou, \emph{{Giant Gravitons - with
  Strings Attached. III.}},
  \href{https://doi.org/10.1088/1126-6708/2008/02/029}{\emph{JHEP} {\bfseries
  02} (2008) 029} [\href{https://arxiv.org/abs/0710.5372}{{\ttfamily
  0710.5372}}].

\bibitem{deMelloKoch:2012sie}
R.~de~Mello~Koch, P.~Diaz and N.~Nokwara, \emph{{Restricted Schur Polynomials
  for Fermions and integrability in the su(2|3) sector}},
  \href{https://doi.org/10.1007/JHEP03(2013)173}{\emph{JHEP} {\bfseries 03}
  (2013) 173} [\href{https://arxiv.org/abs/1212.5935}{{\ttfamily 1212.5935}}].

\bibitem{Mattioli:2016gyl}
P.~Mattioli and S.~Ramgoolam, \emph{{Gauge Invariants and Correlators in
  Flavoured Quiver Gauge Theories}},
  \href{https://doi.org/10.1016/j.nuclphysb.2016.08.021}{\emph{Nucl. Phys. B}
  {\bfseries 911} (2016) 638}
  [\href{https://arxiv.org/abs/1603.04369}{{\ttfamily 1603.04369}}].

\bibitem{Berenstein:2005aa}
D.~Berenstein, \emph{{Large N BPS states and emergent quantum gravity}},
  \href{https://doi.org/10.1088/1126-6708/2006/01/125}{\emph{JHEP} {\bfseries
  01} (2006) 125} [\href{https://arxiv.org/abs/hep-th/0507203}{{\ttfamily
  hep-th/0507203}}].

\bibitem{Berenstein:2004kk}
D.~Berenstein, \emph{{A Toy model for the AdS / CFT correspondence}},
  \href{https://doi.org/10.1088/1126-6708/2004/07/018}{\emph{JHEP} {\bfseries
  07} (2004) 018} [\href{https://arxiv.org/abs/hep-th/0403110}{{\ttfamily
  hep-th/0403110}}].

\bibitem{Lin:2004nb}
H.~Lin, O.~Lunin and J.M.~Maldacena, \emph{{Bubbling AdS space and 1/2 BPS
  geometries}},
  \href{https://doi.org/10.1088/1126-6708/2004/10/025}{\emph{JHEP} {\bfseries
  10} (2004) 025} [\href{https://arxiv.org/abs/hep-th/0409174}{{\ttfamily
  hep-th/0409174}}].

\bibitem{harish1957differential}
Harish-Chandra, \emph{Differential operators on a semisimple lie algebra},
  {\emph{American Journal of Mathematics} (1957) 87}.

\bibitem{Itzykson:1979fi}
C.~Itzykson and J.B.~Zuber, \emph{{The Planar Approximation. 2.}},
  \href{https://doi.org/10.1063/1.524438}{\emph{J. Math. Phys.} {\bfseries 21}
  (1980) 411}.

\bibitem{Zinn-Justin:2002rai}
P.~Zinn-Justin and J.B.~Zuber, \emph{{On some integrals over the U(N) unitary
  group and their large N limit}},
  \href{https://doi.org/10.1088/0305-4470/36/12/318}{\emph{J. Phys. A}
  {\bfseries 36} (2003) 3173}
  [\href{https://arxiv.org/abs/math-ph/0209019}{{\ttfamily math-ph/0209019}}].

\bibitem{Morozov:2009jv}
A.Y.~Morozov, \emph{{Unitary Integrals and Related Matrix Models}},
  \href{https://doi.org/10.1007/s11232-010-0001-y}{\emph{Theor. Math. Phys.}
  {\bfseries 162} (2010) 1} [\href{https://arxiv.org/abs/0906.3518}{{\ttfamily
  0906.3518}}].

\bibitem{Duistermaat:1982vw}
J.J.~Duistermaat and G.J.~Heckman, \emph{{On the Variation in the cohomology of
  the symplectic form of the reduced phase space}},
  \href{https://doi.org/10.1007/BF01399506}{\emph{Invent. Math.} {\bfseries 69}
  (1982) 259}.

\bibitem{Drukker:2000rr}
N.~Drukker and D.J.~Gross, \emph{{An Exact prediction of N=4 SUSYM theory for
  string theory}}, \href{https://doi.org/10.1063/1.1372177}{\emph{J. Math.
  Phys.} {\bfseries 42} (2001) 2896}
  [\href{https://arxiv.org/abs/hep-th/0010274}{{\ttfamily hep-th/0010274}}].

\bibitem{Erickson:2000af}
J.K.~Erickson, G.W.~Semenoff and K.~Zarembo, \emph{{Wilson loops in N=4
  supersymmetric Yang-Mills theory}},
  \href{https://doi.org/10.1016/S0550-3213(00)00300-X}{\emph{Nucl. Phys. B}
  {\bfseries 582} (2000) 155}
  [\href{https://arxiv.org/abs/hep-th/0003055}{{\ttfamily hep-th/0003055}}].

\bibitem{Pestun:2007rz}
V.~Pestun, \emph{{Localization of gauge theory on a four-sphere and
  supersymmetric Wilson loops}},
  \href{https://doi.org/10.1007/s00220-012-1485-0}{\emph{Commun. Math. Phys.}
  {\bfseries 313} (2012) 71} [\href{https://arxiv.org/abs/0712.2824}{{\ttfamily
  0712.2824}}].

\bibitem{Pestun:2016zxk}
V.~Pestun et~al., \emph{{Localization techniques in quantum field theories}},
  \href{https://doi.org/10.1088/1751-8121/aa63c1}{\emph{J. Phys. A} {\bfseries
  50} (2017) 440301} [\href{https://arxiv.org/abs/1608.02952}{{\ttfamily
  1608.02952}}].

\bibitem{Caputa:2013hr}
P.~Caputa, R.~de~Mello~Koch and P.~Diaz, \emph{{A basis for large operators in
  N=4 SYM with orthogonal gauge group}},
  \href{https://doi.org/10.1007/JHEP03(2013)041}{\emph{JHEP} {\bfseries 03}
  (2013) 041} [\href{https://arxiv.org/abs/1301.1560}{{\ttfamily 1301.1560}}].

\bibitem{Caputa:2013vla}
P.~Caputa, R.~de~Mello~Koch and P.~Diaz, \emph{{Operators, Correlators and Free
  Fermions for SO(N) and Sp(N)}},
  \href{https://doi.org/10.1007/JHEP06(2013)018}{\emph{JHEP} {\bfseries 06}
  (2013) 018} [\href{https://arxiv.org/abs/1303.7252}{{\ttfamily 1303.7252}}].

\bibitem{Jackson:1996jb}
A.D.~Jackson, M.K.~Sener and J.J.M.~Verbaarschot, \emph{{Finite volume
  partition functions and Itzykson-Zuber integrals}},
  \href{https://doi.org/10.1016/0370-2693(96)00993-8}{\emph{Phys. Lett. B}
  {\bfseries 387} (1996) 355}
  [\href{https://arxiv.org/abs/hep-th/9605183}{{\ttfamily hep-th/9605183}}].

\bibitem{Byrd:1997uq}
M.~Byrd, \emph{{The Geometry of SU(3)}},
  \href{https://arxiv.org/abs/physics/9708015}{{\ttfamily physics/9708015}}.

\bibitem{Tilma:2002ke}
T.E.~Tilma and G.~Sudarshan, \emph{{Generalized Euler angle parametrization for
  SU(N)}}, \href{https://doi.org/10.1088/0305-4470/35/48/316}{\emph{J. Phys. A}
  {\bfseries 35} (2002) 10467}
  [\href{https://arxiv.org/abs/math-ph/0205016}{{\ttfamily math-ph/0205016}}].

\bibitem{Jiang:2019xdz}
Y.~Jiang, S.~Komatsu and E.~Vescovi, \emph{{Structure constants in $
  \mathcal{N} $ = 4 SYM at finite coupling as worldsheet g-function}},
  \href{https://doi.org/10.1007/JHEP07(2020)037}{\emph{JHEP} {\bfseries 07}
  (2020) 037} [\href{https://arxiv.org/abs/1906.07733}{{\ttfamily
  1906.07733}}].

\bibitem{Jiang:2019zig}
Y.~Jiang, S.~Komatsu and E.~Vescovi, \emph{{Exact Three-Point Functions of
  Determinant Operators in Planar $N=4$ Supersymmetric Yang-Mills Theory}},
  \href{https://doi.org/10.1103/PhysRevLett.123.191601}{\emph{Phys. Rev. Lett.}
  {\bfseries 123} (2019) 191601}
  [\href{https://arxiv.org/abs/1907.11242}{{\ttfamily 1907.11242}}].

\bibitem{Chen:2019kgc}
G.~Chen, R.~De~Mello~Koch, M.~Kim and H.J.R.~Van~Zyl, \emph{{Structure
  constants of heavy operators in ABJM and ABJ theory}},
  \href{https://doi.org/10.1103/PhysRevD.100.086019}{\emph{Phys. Rev. D}
  {\bfseries 100} (2019) 086019}
  [\href{https://arxiv.org/abs/1909.03215}{{\ttfamily 1909.03215}}].

\bibitem{Yang:2021kot}
P.~Yang, Y.~Jiang, S.~Komatsu and J.-B.~Wu, \emph{{D-branes and Orbit
  Average}}, \href{https://doi.org/10.21468/SciPostPhys.12.2.055}{\emph{SciPost
  Phys.} {\bfseries 12} (2022) 055}
  [\href{https://arxiv.org/abs/2103.16580}{{\ttfamily 2103.16580}}].

\bibitem{Budzik:2021fyh}
K.~Budzik and D.~Gaiotto, \emph{{Giant gravitons in twisted holography}},
  \href{https://arxiv.org/abs/2106.14859}{{\ttfamily 2106.14859}}.

\bibitem{Berenstein:2013md}
D.~Berenstein, \emph{{Giant gravitons: a collective coordinate approach}},
  \href{https://doi.org/10.1103/PhysRevD.87.126009}{\emph{Phys. Rev. D}
  {\bfseries 87} (2013) 126009}
  [\href{https://arxiv.org/abs/1301.3519}{{\ttfamily 1301.3519}}].

\bibitem{Gaiotto:2021xce}
D.~Gaiotto and J.H.~Lee, \emph{{The Giant Graviton Expansion}},
  \href{https://arxiv.org/abs/2109.02545}{{\ttfamily 2109.02545}}.

\bibitem{Berenstein:2002ke}
D.~Berenstein, C.P.~Herzog and I.R.~Klebanov, \emph{{Baryon spectra and AdS
  /CFT correspondence}},
  \href{https://doi.org/10.1088/1126-6708/2002/06/047}{\emph{JHEP} {\bfseries
  06} (2002) 047} [\href{https://arxiv.org/abs/hep-th/0202150}{{\ttfamily
  hep-th/0202150}}].

\bibitem{Balasubramanian:2002sa}
V.~Balasubramanian, M.-x.~Huang, T.S.~Levi and A.~Naqvi, \emph{{Open strings
  from N=4 superYang-Mills}},
  \href{https://doi.org/10.1088/1126-6708/2002/08/037}{\emph{JHEP} {\bfseries
  08} (2002) 037} [\href{https://arxiv.org/abs/hep-th/0204196}{{\ttfamily
  hep-th/0204196}}].

\bibitem{Berenstein:2003ah}
D.~Berenstein, \emph{{Shape and holography: Studies of dual operators to giant
  gravitons}},
  \href{https://doi.org/10.1016/j.nuclphysb.2003.10.004}{\emph{Nucl. Phys. B}
  {\bfseries 675} (2003) 179}
  [\href{https://arxiv.org/abs/hep-th/0306090}{{\ttfamily hep-th/0306090}}].

\bibitem{Berenstein:2005vf}
D.~Berenstein and S.E.~Vazquez, \emph{{Integrable open spin chains from giant
  gravitons}}, \href{https://doi.org/10.1088/1126-6708/2005/06/059}{\emph{JHEP}
  {\bfseries 06} (2005) 059}
  [\href{https://arxiv.org/abs/hep-th/0501078}{{\ttfamily hep-th/0501078}}].

\bibitem{Berenstein:2005fa}
D.~Berenstein, D.H.~Correa and S.E.~Vazquez, \emph{{Quantizing open spin chains
  with variable length: An Example from giant gravitons}},
  \href{https://doi.org/10.1103/PhysRevLett.95.191601}{\emph{Phys. Rev. Lett.}
  {\bfseries 95} (2005) 191601}
  [\href{https://arxiv.org/abs/hep-th/0502172}{{\ttfamily hep-th/0502172}}].

\bibitem{Berenstein:2006qk}
D.~Berenstein, D.H.~Correa and S.E.~Vazquez, \emph{{A Study of open strings
  ending on giant gravitons, spin chains and integrability}},
  \href{https://doi.org/10.1088/1126-6708/2006/09/065}{\emph{JHEP} {\bfseries
  09} (2006) 065} [\href{https://arxiv.org/abs/hep-th/0604123}{{\ttfamily
  hep-th/0604123}}].

\bibitem{Berenstein:2013eya}
D.~Berenstein and E.~Dzienkowski, \emph{{Open spin chains for giant gravitons
  and relativity}}, \href{https://doi.org/10.1007/JHEP08(2013)047}{\emph{JHEP}
  {\bfseries 08} (2013) 047} [\href{https://arxiv.org/abs/1305.2394}{{\ttfamily
  1305.2394}}].

\bibitem{Berenstein:2014isa}
D.~Berenstein and E.~Dzienkowski, \emph{{Giant gravitons and the emergence of
  geometric limits in beta-deformations of $ \mathcal{N}=4 $ SYM}},
  \href{https://doi.org/10.1007/JHEP01(2015)126}{\emph{JHEP} {\bfseries 01}
  (2015) 126} [\href{https://arxiv.org/abs/1408.3620}{{\ttfamily 1408.3620}}].

\bibitem{Dzienkowski:2015zba}
E.~Dzienkowski, \emph{{Excited States of Open Strings From $\mathcal{N}=4$
  SYM}}, \href{https://doi.org/10.1007/JHEP12(2015)036}{\emph{JHEP} {\bfseries
  12} (2015) 036} [\href{https://arxiv.org/abs/1507.01595}{{\ttfamily
  1507.01595}}].

\bibitem{Berenstein:2014zxa}
D.~Berenstein, \emph{{On the central charge extension of the $ \mathcal{N}=4 $
  SYM spin chain}}, \href{https://doi.org/10.1007/JHEP05(2015)129}{\emph{JHEP}
  {\bfseries 05} (2015) 129} [\href{https://arxiv.org/abs/1411.5921}{{\ttfamily
  1411.5921}}].

\bibitem{Berenstein:2007wi}
D.~Berenstein, \emph{{Strings on conifolds from strong coupling dynamics, part
  I}}, \href{https://doi.org/10.1088/1126-6708/2008/04/002}{\emph{JHEP}
  {\bfseries 04} (2008) 002} [\href{https://arxiv.org/abs/0710.2086}{{\ttfamily
  0710.2086}}].

\bibitem{Klebanov:2000hb}
I.R.~Klebanov and M.J.~Strassler, \emph{{Supergravity and a confining gauge
  theory: Duality cascades and chi SB resolution of naked singularities}},
  \href{https://doi.org/10.1088/1126-6708/2000/08/052}{\emph{JHEP} {\bfseries
  08} (2000) 052} [\href{https://arxiv.org/abs/hep-th/0007191}{{\ttfamily
  hep-th/0007191}}].

\bibitem{Gopakumar:1998ki}
R.~Gopakumar and C.~Vafa, \emph{{On the gauge theory / geometry
  correspondence}},
  \href{https://doi.org/10.4310/ATMP.1999.v3.n5.a5}{\emph{Adv. Theor. Math.
  Phys.} {\bfseries 3} (1999) 1415}
  [\href{https://arxiv.org/abs/hep-th/9811131}{{\ttfamily hep-th/9811131}}].

\bibitem{Berenstein:2002jq}
D.E.~Berenstein, J.M.~Maldacena and H.S.~Nastase, \emph{{Strings in flat space
  and pp waves from N=4 superYang-Mills}},
  \href{https://doi.org/10.1088/1126-6708/2002/04/013}{\emph{JHEP} {\bfseries
  04} (2002) 013} [\href{https://arxiv.org/abs/hep-th/0202021}{{\ttfamily
  hep-th/0202021}}].

\bibitem{Beisert:2002ff}
N.~Beisert, C.~Kristjansen, J.~Plefka and M.~Staudacher, \emph{{BMN gauge
  theory as a quantum mechanical system}},
  \href{https://doi.org/10.1016/S0370-2693(03)00269-7}{\emph{Phys. Lett. B}
  {\bfseries 558} (2003) 229}
  [\href{https://arxiv.org/abs/hep-th/0212269}{{\ttfamily hep-th/0212269}}].

\bibitem{Berenstein:2004ys}
D.~Berenstein and S.A.~Cherkis, \emph{{Deformations of N=4 SYM and integrable
  spin chain models}},
  \href{https://doi.org/10.1016/j.nuclphysb.2004.09.005}{\emph{Nucl. Phys. B}
  {\bfseries 702} (2004) 49}
  [\href{https://arxiv.org/abs/hep-th/0405215}{{\ttfamily hep-th/0405215}}].

\bibitem{Minahan:2002ve}
J.A.~Minahan and K.~Zarembo, \emph{{The Bethe ansatz for N=4 superYang-Mills}},
  \href{https://doi.org/10.1088/1126-6708/2003/03/013}{\emph{JHEP} {\bfseries
  03} (2003) 013} [\href{https://arxiv.org/abs/hep-th/0212208}{{\ttfamily
  hep-th/0212208}}].

\bibitem{Beisert:2003jj}
N.~Beisert, \emph{{The complete one loop dilatation operator of N=4
  superYang-Mills theory}},
  \href{https://doi.org/10.1016/j.nuclphysb.2003.10.019}{\emph{Nucl. Phys. B}
  {\bfseries 676} (2004) 3}
  [\href{https://arxiv.org/abs/hep-th/0307015}{{\ttfamily hep-th/0307015}}].

\bibitem{Beisert:2010jr}
N.~Beisert et~al., \emph{{Review of AdS/CFT Integrability: An Overview}},
  \href{https://doi.org/10.1007/s11005-011-0529-2}{\emph{Lett. Math. Phys.}
  {\bfseries 99} (2012) 3} [\href{https://arxiv.org/abs/1012.3982}{{\ttfamily
  1012.3982}}].

\bibitem{Berenstein:2020grg}
D.~Berenstein and A.~Holguin, \emph{{Open giant magnons suspended between dual
  giant gravitons in $ \mathcal{N} $ = 4 SYM}},
  \href{https://doi.org/10.1007/JHEP09(2020)019}{\emph{JHEP} {\bfseries 09}
  (2020) 019} [\href{https://arxiv.org/abs/2006.08649}{{\ttfamily
  2006.08649}}].

\bibitem{Berenstein:2020jen}
D.~Berenstein and A.~Holguin, \emph{{Open giant magnons on LLM geometries}},
  \href{https://doi.org/10.1007/JHEP01(2021)080}{\emph{JHEP} {\bfseries 01}
  (2021) 080} [\href{https://arxiv.org/abs/2010.02236}{{\ttfamily
  2010.02236}}].

\bibitem{deMelloKoch:2011ci}
R.~de~Mello~Koch, G.~Kemp and S.~Smith, \emph{{From Large N Nonplanar Anomalous
  Dimensions to Open Spring Theory}},
  \href{https://doi.org/10.1016/j.physletb.2012.04.018}{\emph{Phys. Lett. B}
  {\bfseries 711} (2012) 398}
  [\href{https://arxiv.org/abs/1111.1058}{{\ttfamily 1111.1058}}].

\bibitem{Shatashvili:1992fw}
S.L.~Shatashvili, \emph{{Correlation functions in the Itzykson-Zuber model}},
  \href{https://doi.org/10.1007/BF02097004}{\emph{Commun. Math. Phys.}
  {\bfseries 154} (1993) 421}
  [\href{https://arxiv.org/abs/hep-th/9209083}{{\ttfamily hep-th/9209083}}].

\bibitem{Morozov:1992zb}
A.~Morozov, \emph{{Pair correlator in the Itzykson-Zuber integral}},
  \href{https://doi.org/10.1142/S0217732392002913}{\emph{Mod. Phys. Lett. A}
  {\bfseries 7} (1992) 3503}
  [\href{https://arxiv.org/abs/hep-th/9209074}{{\ttfamily hep-th/9209074}}].

\bibitem{Eynard:2005wg}
B.~Eynard and A.P.~Ferrer, \emph{{2-matrix versus complex matrix model,
  integrals over the unitary group as triangular integrals}},
  \href{https://doi.org/10.1007/s00220-006-1541-8}{\emph{Commun. Math. Phys.}
  {\bfseries 264} (2006) 115}
  [\href{https://arxiv.org/abs/hep-th/0502041}{{\ttfamily hep-th/0502041}}].

\bibitem{deMelloKoch:2020agz}
R.~de~Mello~Koch, J.-H.~Huang, M.~Kim and H.J.R.~Van~Zyl, \emph{{Emergent
  Yang-Mills theory}},
  \href{https://doi.org/10.1007/JHEP10(2020)100}{\emph{JHEP} {\bfseries 10}
  (2020) 100} [\href{https://arxiv.org/abs/2005.02731}{{\ttfamily
  2005.02731}}].

\bibitem{deMelloKoch:2020jmf}
R.~de~Mello~Koch, E.~Gandote and A.L.~Mahu, \emph{{Scrambling in Yang-Mills}},
  \href{https://doi.org/10.1007/JHEP01(2021)058}{\emph{JHEP} {\bfseries 01}
  (2021) 058} [\href{https://arxiv.org/abs/2008.12409}{{\ttfamily
  2008.12409}}].

\bibitem{deMelloKoch:2012ck}
R.~de~Mello~Koch and S.~Ramgoolam, \emph{{A double coset ansatz for
  integrability in AdS/CFT}},
  \href{https://doi.org/10.1007/JHEP06(2012)083}{\emph{JHEP} {\bfseries 06}
  (2012) 083} [\href{https://arxiv.org/abs/1204.2153}{{\ttfamily 1204.2153}}].

\bibitem{deMelloKoch:2017ytj}
R.~de~Mello~Koch and L.~Nkumane, \emph{{From Gauss Graphs to Giants}},
  \href{https://doi.org/10.1007/JHEP02(2018)005}{\emph{JHEP} {\bfseries 02}
  (2018) 005} [\href{https://arxiv.org/abs/1710.09063}{{\ttfamily
  1710.09063}}].

\bibitem{Gomis:2006im}
J.~Gomis and F.~Passerini, \emph{{Wilson Loops as D3-Branes}},
  \href{https://doi.org/10.1088/1126-6708/2007/01/097}{\emph{JHEP} {\bfseries
  01} (2007) 097} [\href{https://arxiv.org/abs/hep-th/0612022}{{\ttfamily
  hep-th/0612022}}].

\bibitem{Berenstein:2004hw}
D.~Berenstein, \emph{{A Matrix model for a quantum Hall droplet with manifest
  particle-hole symmetry}},
  \href{https://doi.org/10.1103/PhysRevD.71.085001}{\emph{Phys. Rev. D}
  {\bfseries 71} (2005) 085001}
  [\href{https://arxiv.org/abs/hep-th/0409115}{{\ttfamily hep-th/0409115}}].

\bibitem{Berenstein:2019esh}
D.~Berenstein and R.~de~Mello~Koch, \emph{{Gauged fermionic matrix quantum
  mechanics}}, \href{https://doi.org/10.1007/JHEP03(2019)185}{\emph{JHEP}
  {\bfseries 03} (2019) 185}
  [\href{https://arxiv.org/abs/1903.01628}{{\ttfamily 1903.01628}}].

\bibitem{Berenstein:2017abm}
D.~Berenstein and A.~Miller, \emph{{Superposition induced topology changes in
  quantum gravity}}, \href{https://doi.org/10.1007/JHEP11(2017)121}{\emph{JHEP}
  {\bfseries 11} (2017) 121}
  [\href{https://arxiv.org/abs/1702.03011}{{\ttfamily 1702.03011}}].

\bibitem{Lin:2017dnz}
H.~Lin and K.~Zeng, \emph{{Detecting topology change via correlations and
  entanglement from gauge/gravity correspondence}},
  \href{https://doi.org/10.1063/1.4986985}{\emph{J. Math. Phys.} {\bfseries 59}
  (2018) 032301} [\href{https://arxiv.org/abs/1705.10776}{{\ttfamily
  1705.10776}}].

\bibitem{Pestun:2009nn}
V.~Pestun, \emph{{Localization of the four-dimensional N=4 SYM to a two-sphere
  and 1/8 BPS Wilson loops}},
  \href{https://doi.org/10.1007/JHEP12(2012)067}{\emph{JHEP} {\bfseries 12}
  (2012) 067} [\href{https://arxiv.org/abs/0906.0638}{{\ttfamily 0906.0638}}].

\bibitem{Berenstein:2006yy}
D.~Berenstein and R.~Cotta, \emph{{Aspects of emergent geometry in the AdS/CFT
  context}}, \href{https://doi.org/10.1103/PhysRevD.74.026006}{\emph{Phys. Rev.
  D} {\bfseries 74} (2006) 026006}
  [\href{https://arxiv.org/abs/hep-th/0605220}{{\ttfamily hep-th/0605220}}].

\bibitem{Berenstein:2007kq}
D.E.~Berenstein and S.A.~Hartnoll, \emph{{Strings on conifolds from strong
  coupling dynamics: Quantitative results}},
  \href{https://doi.org/10.1088/1126-6708/2008/03/072}{\emph{JHEP} {\bfseries
  03} (2008) 072} [\href{https://arxiv.org/abs/0711.3026}{{\ttfamily
  0711.3026}}].

\bibitem{Biswas:2006tj}
I.~Biswas, D.~Gaiotto, S.~Lahiri and S.~Minwalla, \emph{{Supersymmetric states
  of N=4 Yang-Mills from giant gravitons}},
  \href{https://doi.org/10.1088/1126-6708/2007/12/006}{\emph{JHEP} {\bfseries
  12} (2007) 006} [\href{https://arxiv.org/abs/hep-th/0606087}{{\ttfamily
  hep-th/0606087}}].

\bibitem{Pasukonis:2010rv}
J.~Pasukonis and S.~Ramgoolam, \emph{{From counting to construction of BPS
  states in N=4 SYM}},
  \href{https://doi.org/10.1007/JHEP02(2011)078}{\emph{JHEP} {\bfseries 02}
  (2011) 078} [\href{https://arxiv.org/abs/1010.1683}{{\ttfamily 1010.1683}}].

\bibitem{Lewis-Brown:2020nmg}
C.~Lewis-Brown and S.~Ramgoolam, \emph{{Quarter-BPS states, multi-symmetric
  functions and set partitions}},
  \href{https://doi.org/10.1007/JHEP03(2021)153}{\emph{JHEP} {\bfseries 03}
  (2021) 153} [\href{https://arxiv.org/abs/2007.01734}{{\ttfamily
  2007.01734}}].

\bibitem{Aharony:2008ug}
O.~Aharony, O.~Bergman, D.L.~Jafferis and J.~Maldacena, \emph{{N=6
  superconformal Chern-Simons-matter theories, M2-branes and their gravity
  duals}}, \href{https://doi.org/10.1088/1126-6708/2008/10/091}{\emph{JHEP}
  {\bfseries 10} (2008) 091} [\href{https://arxiv.org/abs/0806.1218}{{\ttfamily
  0806.1218}}].

\bibitem{Drukker:2009hy}
N.~Drukker and D.~Trancanelli, \emph{{A Supermatrix model for N=6 super
  Chern-Simons-matter theory}},
  \href{https://doi.org/10.1007/JHEP02(2010)058}{\emph{JHEP} {\bfseries 02}
  (2010) 058} [\href{https://arxiv.org/abs/0912.3006}{{\ttfamily 0912.3006}}].

\bibitem{Drukker:2010nc}
N.~Drukker, M.~Marino and P.~Putrov, \emph{{From weak to strong coupling in
  ABJM theory}}, \href{https://doi.org/10.1007/s00220-011-1253-6}{\emph{Commun.
  Math. Phys.} {\bfseries 306} (2011) 511}
  [\href{https://arxiv.org/abs/1007.3837}{{\ttfamily 1007.3837}}].

\bibitem{Minahan:2008hf}
J.A.~Minahan and K.~Zarembo, \emph{{The Bethe ansatz for superconformal
  Chern-Simons}},
  \href{https://doi.org/10.1088/1126-6708/2008/09/040}{\emph{JHEP} {\bfseries
  09} (2008) 040} [\href{https://arxiv.org/abs/0806.3951}{{\ttfamily
  0806.3951}}].

\bibitem{Beem:2013sza}
C.~Beem, M.~Lemos, P.~Liendo, W.~Peelaers, L.~Rastelli and B.C.~van Rees,
  \emph{{Infinite Chiral Symmetry in Four Dimensions}},
  \href{https://doi.org/10.1007/s00220-014-2272-x}{\emph{Commun. Math. Phys.}
  {\bfseries 336} (2015) 1359}
  [\href{https://arxiv.org/abs/1312.5344}{{\ttfamily 1312.5344}}].

\end{thebibliography}\endgroup

\end{document}